\documentclass[showpacs,aps, twocolumn,nofootinbib]{revtex4-2}
\usepackage{epsfig}
\usepackage{graphicx}
\usepackage{amsmath,amssymb,amsfonts}
\usepackage{array}
\usepackage{url}
\usepackage{xcolor}
\usepackage[colorlinks=true,linkcolor=purple, citecolor = blue]{hyperref}
\usepackage{multirow}
\usepackage{float}
\usepackage{lineno}
\usepackage{xspace}
\usepackage{ulem}
\usepackage{lipsum} 
\usepackage{footnote}

\begin{document}

\title{Spin alignment of Quarkonia: A Possible Probe of Deconfined QCD matter in Pb+Pb Collisions at $\sqrt{s_{\rm NN}} = 5.02$ TeV}

\author{Bhagyarathi Sahoo}
\email{Bhagyarathi.Sahoo@cern.ch}
\author{Captain R. Singh}
\email{captainriturajsingh@gmail.com}
\author{Raghunath Sahoo}
\email{Raghunath.Sahoo@cern.ch (Corresponding author)}
\affiliation{Department of Physics, Indian Institute of Technology Indore, Simrol, Indore 453552, India}

\begin{abstract}

In this study, we investigate the influence of deconfined QCD matter on quarkonium spin alignment in ultra-relativistic heavy-ion collisions. We estimate the spin alignment of charmonium ($J/\psi$, and $\psi$(2S)) and bottomonium ($\Upsilon$(1S), and $\Upsilon$(2S)) states for Pb+Pb collisions at $\sqrt{s_{\rm NN}} = 5.02$ TeV as a function of transverse momentum by calculating the energy eigenvalues in a thermal rotating medium. We solve the Schrödinger equation with a medium-modified color-singlet potential, considering the coupling of spin with vorticity and magnetic field. Furthermore, we evaluate the effect of medium temperature, vorticity, magnetic field, and momentum-space anisotropy on the elements of the spin density matrix. Our findings reveal that vorticity increases the spin alignment, while the magnetic fields and anisotropy modify the observables in a state-dependent manner. These findings deepen our understanding of quarkonium spin alignment in an anisotropic magneto-vortical thermal medium, shedding light on spin transport phenomena in heavy-ion collisions.

\end{abstract}
\date{\today}
\maketitle

\section{Introduction}
\label{intro}
Conventionally, the thermalized deconfined state of strongly interacting partons known as quark-gluon plasma (QGP) has 
been investigated in heavy-ion collisions (HIC), considering the $pp$ collisions as a baseline. It was believed that the QGP 
existence in $pp$ collisions is next to impossible, as it lacks the necessary conditions for QGP to exist. However, recent 
studies suggest that at high multiplicity $pp$ collisions have shown behavior similar to heavy-ion collisions, such as 
collective flow, strangeness enhancement, etc.~\cite{ALICE:2016fzo, CMS:2016fnw}. Moreover, few studies advocate that similar 
phenomena may arise in perturbative QCD-based models which do not consider the medium formation into 
account~\cite{OrtizVelasquez:2013ofg, Zhang:2021phk, Prasad:2024gqq}. These investigations raise questions about the current QGP signatures and highlight the necessity for 
a comprehensive study of QGP signals and collective behavior. In the quest for such a baseline-independent probe, spin polarization comes into the picture and can have implications for understanding the hot QCD matter. Recent studies present that the dilepton polarization could be a probe of thermalization of QGP~\cite{Coquet:2023wjk,Wu:2024vyc}. However, there are various sources by which particles can get polarized in ultra-relativistic nuclear 
collisions. Such as the vorticity field created due to the initial orbital angular momentum (OAM) in the peripheral 
collisions can polarize the quarks, and such polarization is transferred to the final state hadrons. The magnetic field 
produced by the charged spectators in such collisions interacts with the spin of the quarks and plays a 
vital role in heavy-quark polarization~\cite{Dey:2025ail}. Additionally, the particle production mechanisms constrain the polarization 
of the hadrons in hadronic and nuclear collisions~\cite{Sahoo:2023oid}. Determining the precise contribution of each source to the spin 
polarization observables in ultra-relativistic collisions requires comprehensive theoretical studies with corresponding experimental 
observations.\\

The experimental measurements of global and local spin polarization of $\Lambda$ and $\bar{\Lambda}$ 
hyperons~\cite{STAR:2017ckg,STAR:2018gyt, STAR:2019erd,STAR:2023eck,STAR:2020xbm, ALICE:2021pzu, ALICE:2019onw} 
have been qualitatively explained using the hydrodynamic and transport models~\cite{Xie:2017upb,Ryu:2021lnx,Fu:2020oxj,Wei:2018zfb,Becattini:2021iol,Sahoo:2024egx}. Apart from the spin polarization of baryons, the so-called dual observables, i.e., the spin alignment of vector mesons, have been measured for $K^{*0}$, $\phi$, $\rm D^{*+}$, and $K_{S}^{0}$~\cite{STAR:2008lcm, STAR:2022fan, ALICE:2019aid, ALICE:2022byg, ALICE:2025cdf}. Furthermore, the spin alignment of higher 
resonance vector mesons such as $J/\psi$, $\psi$(2S), $\Upsilon$(1S), $\Upsilon$(2S), and $\Upsilon(3S)$ states have been measured in  
hadronic~\cite{STAR:2013iae, STAR:2020igu, CMS:2013gbz, CMS:2012bpf, ALICE:2011gej, ALICE:2018crw, LHCb:2013izl, LHCb:2014brf} and 
nuclear~\cite{ALICE:2020iev, ALICE:2022dyy} collisions. The measurements of spin polarization of baryons and the spin alignment of vector mesons in nuclear collisions indicate the creation of huge vorticity in the system. Thus, it is important to understand the various possible sources of vorticity formation in heavy-ion collisions.\\

Besides OAM, there could be several other potential sources of vorticity formation in the system. For instance, jet-like 
fluctuations in the expanding fireball can create smoke-loop type vortices around fast-moving particles~\cite{Betz:2007kg}. 
The inhomogeneous expansion of the fireball can also generate vorticity~\cite{Xia:2018tes, Wei:2018zfb}. The finite viscosity present in the system can produce vorticity~\cite{Sahoo:2023xnu}.
Furthermore, the magnetized QCD matter induces a finite rotation in the system due to the Einstein-de Haas effect~\cite{Einstein:1915}. The strength of the vorticity field ($\omega$ $\approx$ 0.02 fm$^{-1}$ $\approx$ 10$^{22}$ s$^{-1}$) produced in heavy-ion collisions can be estimated from the average global hyperon polarization measurements of $\Lambda$ and $\bar{\Lambda}$ hyperons using a statistical thermal model~\cite{Becattini:2016gvu}. Recently, it has been proposed that the strength of the vorticity field can be estimated from the directed flow measurements of hadrons~\cite{Jiang:2023vxp}. Accompanying the vorticity field, a substantial magnetic field ($eB \approx m_{\pi}^{2} \approx 10^{18} $ G) is produced due to the 
charged spectators in peripheral heavy ion collisions. The strength of the magnetic field may reach up to the order of 0.1$m_{\pi}^{2}$, 
$m_{\pi}^{2}$, and $15 m_{\pi}^{2}$ for SPS, RHIC, and LHC energies, respectively~\cite{Skokov:2009qp}. Both the vorticity field, and magnetic field provide new insights into heavy-ion collisions dynamics and significantly influence QCD thermodynamics~\cite{Sahoo:2023vkw, Pradhan:2023rvf}, transport properties~\cite{Aung:2023pjf, Dwibedi:2023akm}, hydrodynamic medium evolution~\cite{Sahoo:2023xnu}, QCD phase transition~\cite{Pradhan:2023rvf}, and spin polarization of hadrons~\cite{Xie:2017upb, Ryu:2021lnx, Fu:2020oxj, Wei:2018zfb, Becattini:2021iol, Sahoo:2024egx, Wu:2023dfz, Han:2017hdi, Xu:2022hql}. Furthermore, 
due to the rapid longitudinal expansion of the fireball created in heavy-ion collisions, a high degree of momentum-space 
anisotropy is produced in the QGP medium in its local rest frame~\cite{Dong:2021gnb}. 
The study of anisotropy has been a subject of investigation in various contexts~\cite{Romatschke:2006bb, Schenke:2006yp, 
Mauricio:2007vz, Dumitru:2007rp, Baier:2008js, Dumitru:2007hy, Carrington:2008sp}. 
Here, we attempted to observe the impact of medium anisotropy on the quarkonia spin alignment in heavy-ion collisions. In this 
article,  anisotropy is introduced at the level of the medium-modified potential through Debye mass ($m_{\rm D}$).\\

In ultra-relativistic heavy-ion collisions, heavy quarks are produced in the initial hard scattering, and therefore, these are considered relatively cleaner probes to investigate the early-stage properties of the hot and dense deconfined QCD  medium. It can be expected that the effect of the initially produced vorticity field, magnetic field, and momentum space anisotropy strongly affects the heavy quarks compared to light quarks. Therefore, with an intense investigation of heavy-flavor hadrons at the hadronization stage, one may extract precise information about the initially produced vorticity, magnetic field, and momentum space anisotropy, and can quantify how much these affect the heavy-flavor dynamics. \\

In this study, we explore these effects on the spin alignment of different quarkonium states using the spin-density matrix elements, which can be measured in the experiment. The spin alignment of vector mesons has been studied by measuring the elements of a 3 $\times$ 3 Hermitian spin density matrix ($\rho_{m,m^{'}}$), where $ m $ and $m^{'}$ label the spin component along the quantization axis. The diagonal elements  $\rho_{11}$, $\rho_{00}$, and $\rho_{-1-1}$ define the probabilities of the spin component of vector mesons along the quantization axis. Out of these three diagonal elements, 
$\rho_{00}$ is independent and can be measured in experiments in two-body decays to pseudoscalar mesons (for example $\phi$ $\to$ $K^{+}K^{-}$). For quarkonia decay such as $J/\psi \to \mu^+ \mu^-$ (or $e^{+}e^{-}$), $\rho_{00}$ is determined from the  angular distribution of decay leptons in the quarkonium rest frame, which is given by~\cite{Sahoo:2023oid};

\begin{equation}
\frac{dN}{d\Omega} \propto \left(1 + \lambda_{\theta} \cos^{2}\theta 
+ \lambda_{\phi} \sin^{2}\theta \cos 2\phi 
+ \lambda_{\theta\phi} \sin 2\theta \cos \phi \right)
\end{equation}

where $\theta$ and $\phi$ are the polar and azimuthal angles of the decay products, while $\lambda_{\theta}$, $\lambda_{\phi}$ and $\lambda_{\theta\phi}$ are the polarization parameters that quantify the anisotropy of the decay distribution. The parameter $\rho_{00}$, which quantifies the probability of the vector meson being in a spin-zero projection along the quantization 
axis, is connected to the polar anisotropy parameter $\lambda_{\theta}$ by the relation,

\begin{equation}
\rho_{00} = \frac{1 - \lambda_{\theta}}{3 + \lambda_{\theta}}
\end{equation}

In the absence of spin alignment, all three spin states of vector mesons 1, 0, and -1 are equally probable, and thus $\rho_{00}$ = 1/3. The deviation of $\rho_{00}$ from 1/3 (i.e. $\rho_{00} - 1/3$) determines the degree of alignment of vector meson. A value $\rho_{00} < 1/3$ indicates a transverse alignment, whereas $\rho_{00} > 1/3$ corresponds to a longitudinal alignment with respect to the chosen quantization axis. A precise determination of $\rho_{00}$ and its associated polarization parameters in different reference frames, such as the helicity, Collins–Soper, and event-plane frames, provides key insights into the spin dynamics of quarkonia and the influence of the surrounding medium in relativistic heavy-ion collisions.\\

On the other hand, the off-diagonal elements of the spin density matrix play a critical role in understanding the spin alignment of quarkonium states in the QGP medium. The diagonal component $\rho_{00}$ is commonly used to quantify spin alignment, indicating the probability of finding a vector meson in a spin-0 projection along a chosen quantization axis, while the off-diagonal elements such as $\rho_{1,-1}$, $\rho_{-1,0}$, and $\rho_{1,0}$ provide information about the quantum coherence and interference between different spin states. The non-zero off-diagonal elements imply that the system is in a superposition of spin states rather than a statistical mixture, thus emphasizing the partial coherence and entanglement within the spin degrees of freedom. These coherence terms are sensitive to the microscopic structure and dynamics of the medium. For instance, in a locally rotating QGP, vorticity gradients may induce spin-orbit couplings that lead to position-dependent polarization, referred to as local spin alignment. This effect cannot be captured solely by $\rho_{00}$, which represents global polarization. Instead, the angular modulations introduced by off-diagonal terms such as $\rho_{1,-1}$ and $\rho_{1,0}$ become crucial in distinguishing local from global alignment. Specifically, $\rho_{1,-1}$ contributes to azimuthal anisotropies in the decay distributions of vector mesons, while $\rho_{-1,0}$ and $\rho_{1,0}$ encode spin correlations along transverse directions. They may also provide valuable insights into medium-induced spin correlations, the local spin structure, and non-equilibrium QCD dynamics. The measurement of the real part of the off-diagonal elements of the spin density matrix such as $\rho_{1,-1}$, the linear combination of $\rho_{-1,0}$, and $\rho_{0,1}$ is embedded in experimental analysis through the angular distribution coefficients $\lambda_\phi$, and $\lambda_{\theta\phi}$. Through experimental fits to such distributions, one can extract these elements and obtain a more complete description of the spin polarization. \\

Various phenomenological approaches have been developed gradually to describe the spin alignment phenomena of vector 
mesons~\cite{Liang:2004xn, Yang:2017sdk, Sheng:2020ghv, Sheng:2022wsy, Xia:2020tyd, Lv:2024uev, Gao:2021rom, Sheng:2019kmk, Goncalves:2021ziy, Goncalves:2024xzo, DeMoura:2023jzz, Sahoo:2024yud, Xu:2024kdh, Kumar:2023ghs, Fu:2023qht, Zhao:2024ipr, Sheng:2024kgg, Chen:2024afy, Muller:2021hpe}. The global quark polarization leading to the spin alignment of vector mesons was first introduced in Ref.~\cite{Liang:2004xn}. A quark coalescence model attempts to explain the spin alignment of vector mesons in Refs.~\cite{Yang:2017sdk, Sheng:2020ghv}. Apart from the global spin alignment, the local spin alignment induced by the local vorticity is discussed in Ref.~\cite{Xia:2020tyd}. The authors have proposed that the measurement of $\rho_{00}$ observable as a function of the azimuthal angle of the particle’s transverse momentum with respect to the reaction plane, off-diagonal elements of the spin-density matrix $\rho_{m,m^{'}}$, and $<\rho_{00}>$ with respect to different event planes can be used to separate the local spin alignment of vector mesons from the global one. In Ref.~\cite{Goncalves:2021ziy, Goncalves:2024xzo, DeMoura:2023jzz}, the authors have argued that the spin alignment of vector mesons can serve as a probe of spin-hydrodynamics and freeze-out with a coalescence model. The spin alignment of vector mesons is explained with the anisotropies of local field correlation (or fluctuation of vector meson force field)~\cite{Sheng:2022wsy, Lv:2024uev}, 
glasma field~\cite{Kumar:2023ghs, Muller:2021hpe}, light front quarks~\cite{Fu:2023qht}, local spin density fluctuations~\cite{Xu:2024kdh}, holographic method \cite{Zhao:2024ipr, Sheng:2024kgg}, hadronization mechanisms~\cite{Sahoo:2024yud, Chen:2024afy}, etc.\\

In this study, we explore the spin alignment of various charmonium states; such as $J/\psi$, and $\psi$(2S), as well as bottomonium states; such as $\Upsilon$(1S), and $\Upsilon$(2S) in Pb+Pb collisions at $\sqrt{s_{\rm NN}} = 5.02$ TeV obtained at mid-rapidity as a function of transverse momentum ($p_{\rm T}$). We investigate the effect of medium rotation or vorticity ($\omega$), magnetic fields ($eB$), and momentum space anisotropy ($\xi$) on the spin alignment observables $\rho_{00}$. We estimate both the diagonal and off-diagonal elements of the spin-density matrix for quarkonia in a thermal rotating medium. To obtain these matrix elements, we solved the Schrödinger equation numerically and obtained the energy eigenvalues using a medium-modified color octet potential model for quarkonia. We extend the standard formalism of solving quarkonium states in a rotating frame to account for the external magnetic field and medium anisotropy. The energy eigenvalues depend upon the coupling of spin with rotation and the magnetic field. \\

The present work extends the study of quarkonium spin alignment in relativistic heavy-ion collisions beyond the framework of Ref.~\cite{DeMoura:2023jzz} in several important aspects. Unlike Ref.~\cite{DeMoura:2023jzz}, which employs a vacuum quarkonium potential without incorporating medium evolution, we utilize a realistic medium-modified quarkonium potential to describe in-medium quarkonium properties. In addition, we implement a second-order viscous hydrodynamic evolution of the vortical QGP medium, allowing us to explicitly study the time evolution of vorticity and its impact on quarkonium spin alignment. Furthermore, our analysis is performed under experimentally relevant conditions for Pb+Pb collisions at $\sqrt{s_{\rm NN}} =$ 5.02 TeV, including centrality and transverse-momentum dependence of both diagonal and off-diagonal spin-density matrix elements. In addition, we investigate the effects of magnetic fields and medium momentum anisotropy on quarkonium spin alignment. These ingredients provide a more realistic and comprehensive framework for connecting the microscopic spin dynamics of quarkonia with the evolving QGP medium and facilitate direct comparison with current and future experimental measurements.\\

This paper is organized as follows: Section \ref{formulation} provides a detailed description of the formalism used in this study, Section \ref{res} discusses the results, and Section \ref{sum} offers a summary of our findings.

\section{Formulation}
\label{formulation}
In the present study, we assume the quarkonium states as quantum particles, and it is characterized by the quantum wavefunction. The evolution of quarkonium states is described using the time-independent Schr\"{o}dinger equation (SE) with a temperature-dependent complex heavy-quark potential. This formalism has been widely used to describe the evolution, production, and suppression of various quarkonium states in heavy-ion and small collision systems at various center-of-mass energies~\cite{ Singh:2015eta, Ganesh:2016kug, Singh:2018wdt, Hatwar:2020esf, Singh:2021evv, Singh:2025xrd}.  In this work, we use the standard developed formalism and investigate the spin-alignment of quarkonium states by incorporating the spin-vorticity and spin-magnetic field coupling in a thermal rotating medium.\\ 

Now, before we explore the quarkonium spin alignment mechanism within the QGP medium, it is necessary to first examine how the quarkonium Hamiltonian is modified in the medium in the presence of vorticity, temperature, and magnetic field. It is well established that the Hamiltonian of a charged particle with charge $q$, mass $m$, and spin $S$, moving in a rotating medium with angular velocity $\omega$ and subjected to a magnetic field $B$, can be expressed as~\cite{Anan, Gordan, Griffth};

\begin{align}
\mathcal{H} = & \; \frac{1}{2m}\left(\boldsymbol{p}  - q\boldsymbol{A} - m\boldsymbol{\omega}\times\mathbf{r} \right)^2   - \frac{m}{2}(\boldsymbol{\omega}\times\mathbf{r})^2 \nonumber \\
&-\boldsymbol{\mu} \cdot\mathbf{B} -\boldsymbol{\omega}\cdot\mathbf{S}+ V(\boldsymbol{r})
\label{RotH}
\end{align}

Here, $\boldsymbol{\omega} \cdot \mathbf{S}$ term arises from the coupling between the rotation of the system and the spin of the particle, while the term $\boldsymbol{\mu} \cdot \mathbf{B}$ results from the coupling between the magnetic field and the spin magnetic dipole moment. The $\boldsymbol{A} = -\frac{1}{2} (\bold{r} \times \bold{B})$, denotes the magnetic vector potential, and $\boldsymbol{\mu}$ stands for the magnetic moment associated with the spin of a particle with charge $q$ and mass $m$. The magnetic moment is defined as $\boldsymbol{\mu} = g\frac{q}{2m}\boldsymbol{S}$, where $g$ is the Land'e $g$-factor.\\

Now, for a two-body system like  quarkonia in QGP medium, Eq.~\ref{RotH} can be written as;

\begin{align}
&\mathcal{H} = \sum_{i=1,2}\left[\frac{1}{2m_i}(\boldsymbol{p}_i-q_{i} \boldsymbol{A_i} - m_i\boldsymbol{\omega}_i\times\mathbf{r}_i)^2 \right. \nonumber \\
& \left. - \frac{m_{i}}{2}(\boldsymbol{\omega_{i}}\times\mathbf{r}_{i})^2 - \boldsymbol{\mu_i} \cdot\mathbf{B_i} - \boldsymbol{\omega_{i}}\cdot\mathbf{S_i}\right]  + V(\lvert \mathbf{r_1} - \mathbf{r_2} \rvert)
\label{RotH2}
\end{align}	
In this study, the index $i$ runs over 1 and 2, representing the heavy quark and antiquark, respectively. Employing the standard reduction procedure of a two-body system to an effective one-body description in the center-of-mass frame and introducing the corresponding reduced coordinate system, the dynamics of the quarkonia can be expressed through the following relation,

\begin{equation}
\left\{
\begin{array}{lcl}
\mathbf{P} = \mathbf{p_1} + \mathbf{p_2},  & &
\mathbf{p} = m_{\mu} \left(\dfrac{\mathbf{p_1}}{m_1} - \dfrac{\mathbf{p_2}}{m_2}\right) \\[3mm]
m_{\mu} = \dfrac{m_1\;m_2}{m_1+m_2},  &\quad & \\[3mm]
\mathbf{r_1} = \mathbf{R} +\dfrac{m_2}{m_1 + m_2}\mathbf{r}, & &
\mathbf{r_2} = \mathbf{R} -\dfrac{m_1}{m_1 + m_2}\mathbf{r}
\end{array}
\right.
\label{reduced}
\end{equation}
where, $m_{\mu}$ = $m_{Q}/2$ represents the reduced mass in the center-of-mass frame of the heavy-quark pair, with the charm quark mass $m_{c} = 1.27$ GeV and bottom quark mass $m_{b} = 4.18$ GeV. 
We assume that the quark and anti-quark pair forming the quarkonium states exist within the same vortical and magnetic fields. Thus, we have $\boldsymbol{\omega}_1 = \boldsymbol{\omega}_2 = \boldsymbol{\omega}$ and $\boldsymbol{B}_1 = \boldsymbol{B}_2 = \boldsymbol{B}$, and following this, the Hamiltonian can be rewritten as,
\begin{align}    
\mathcal{H} =&\; \frac{P^2}{2M} + \frac{p^2}{2m_{\mu}} - \mathbf{P}\cdot(\boldsymbol{\omega}\times\mathbf{R}) - \mathbf{p}\cdot(\boldsymbol{\omega}\times\mathbf{r})\nonumber \\
&- (\boldsymbol{\mu_1} + \boldsymbol{\mu_2} )\cdot \mathbf{B} - \boldsymbol{\omega}\cdot(\mathbf{S}_{1}+\mathbf{S}_{2})+ V(\lvert \mathbf{r} \rvert) 
\label{H2}	
\end{align}

Since we are interested in the reduced coordinate, we obtain

\begin{equation}
\mathcal{H} = \frac{p^2}{2m_{\mu}} - \mathbf{p}\cdot(\boldsymbol{\omega}\times\mathbf{r}) - (\boldsymbol{\mu_1} + \boldsymbol{\mu_2} )\cdot \mathbf{B} - \boldsymbol{\omega}\cdot(\mathbf{S}_{1}+\mathbf{S}_{2}) + V(\lvert \mathbf{r} \rvert) 
\label{H3}
\end{equation}
Assuming the magnetic field $\boldsymbol{B}$ and the rotation $\boldsymbol{\omega}$  along the z direction, such that $ \boldsymbol{B} = B \hat{z}$ and $ \boldsymbol{\omega} = \omega \hat{z}$, the  Eq.~\ref{H3}, can be written as;

\begin{equation}
\mathcal{H} = \frac{p^2}{2m_{\mu}} - \mathbf{p}\cdot(\boldsymbol{\omega}\times\mathbf{r}) - \frac{qB}{2m}(S_{1z} - S_{2z})   - \omega (S_{1z}+S_{2z}) + V(\lvert \mathbf{r} \rvert) 
\label{H4}
\end{equation}

Writing $\mathbf{p} = - i \boldsymbol{\nabla}$, $\mathcal{H} = i\frac{\partial}{\partial t}$, $L_z = -i \frac{\partial}{\partial \phi}$, $S_{1z}+S_{2z}
= S_{z}$, 
we can rewrite the Eq.~\ref{H4} as
\begin{equation}
i\frac{\partial}{\partial t}= -\frac{1}{2m_{\mu}}\boldsymbol{\nabla}^2 - \omega (L_z + S_z) - \frac{qB}{2m}(S_{1z} - S_{2z})  + V(\lvert \mathbf{r} \rvert) 
\label{NonIn}	
\end{equation}

Expressing the Laplacian operator in spherical coordinates and following the separation of variables method, the radial part of the Schr\"{o}dinger equation can be written as;

\begin{align}
\frac{1}{r^2}\frac{d}{dr}\left(r^2\frac{dR(r)}{dr}\right) + 2m_{\mu} \left[(E-V(r))-\frac{l(l+1)}{2 m_{\mu} r^2} \right.\nonumber \\
\left. - \omega  (L_z + S_z) - \frac{qB}{m_{\mu}} (S_{1z} - S_{2z}) \right]R(r)=0	
\label{shrr1}
\end{align}
To solve the Schr\"{o}dinger equation in the presence of rotation, we define $\omega$ in terms of the conserved circulation $C$,

\begin{equation}
C = \oint \vec{v}\cdot \vec{dl}
\label{C}
\end{equation}
Applying the fundamental theorem for curl, which goes by the special name of Stokes' theorem, states that the line integral can be converted into a surface integral. Thus, Eq.~\ref{C} can be written as;
\begin{equation}
C =  \oint \vec{\nabla} \times \vec{v}  \cdot \vec{dS} = 2 \omega \pi r^2 
\label{C2}
\end{equation}
where, $\vec{\omega} = \frac{1}{2} \vec{\nabla} \times \vec{v}$ is the non-relativistic vorticity. Here, we assume the rotation to be classical and can be related to the hydrodynamic vorticity, such as the thermal vorticity, temperature vorticity, and enthalpy vorticity, etc.\\ 

Substituting the value of $C$ in Eq.~\ref{shrr1}, we obtain

\begin{align}
\frac{1}{r^2}\frac{d}{dr}\left(r^2\frac{dR(r)}{dr}\right) + 2m_{\mu} \left[(E-V(r))-\frac{l(l+1)}{2 m_{\mu} r^2} \right. \nonumber \\
\left. - \frac{m_{j}C}{2\pi r^{2}} - \frac{qB}{m_{\mu}} \right]R(r)=0	
\label{shrr2}
\end{align}

The color singlet potential for  $c-\bar{c}$ bound state in the QGP as follows~\cite{nendzig};

\begin{align}
V(r,&m_{\rm D}) = \frac{\sigma}{m_{\rm D}}(1 - e^{-m_{\rm D}\,r}) - \alpha_{\rm eff} \left ( m_{\rm D}
+ \frac{e^{-m_{\rm D}\,r}}{r} \right )\nonumber \\ &-  i\alpha_{\rm eff} T_{\rm eff} \int_0^\infty
\frac{2\,z\,dz}{(1+z^2)^2} \left ( 1 - \frac{\sin(m_{\rm D}\,r\,z)}{m_{\rm D}\,r\,z} \right)
\label{potential}
\end{align}

Where the first and second terms on the right-hand side are the string term and the Coulombic term, respectively. The third term on the right-hand side is the imaginary part of the heavy-quark potential. In the above Eq.~\ref{potential} :

\begin{itemize}
\item $\sigma$ is the string tension constant between $c\bar{c}$ bound
state, and is given as  $\sigma = 0.192$ GeV$^2$.

\item $T_{\rm eff}$ is the effective temperature of quarkonia in the medium, arises due to the Relativistic Doppler Shift (RDS), and it is given as~\cite{Singh:2018wdt};

\begin{equation} 
T_{\rm eff}(p_{T},\tau,  b) = \frac{T (\tau, b)\sqrt{1 -
|v_{r}|^{2}}}{2\;|v_{r}|}\;\ln\Bigg[\frac{1 + |v_{r}|}{1 - |v_{r}|}\Bigg]
\label{Teff}
\end{equation}

Here, T($\tau$, $b$) is the evolving medium temperature depending on evolution time ($\tau$) and impact parameter ($b$) between the colliding ions. The $v_{r}$ is the relative velocity between the medium ($v_m$) and quarkonia ($v_{Q}$). The thermal velocity ($v_m$) of the medium is obtained using the Maxwell-J\"uttner velocity distribution~\cite{Singh:2025xrd}. The velocity of the quarkonium states is obtained as, $v_{Q}$ = $p_{T}/E_{T}$, here $p_{T}$ is transverse momentum of the quarkonia, $E_T = \sqrt{p_{T}^{2} + M^{2}_{nl}}$ is its transverse energy, and $M_{nl}$ is the mass of the quarkonium state. 
It is important to note that Eq.~\ref{Teff} is valid under the limit $v_r \ll 1$. For a more detailed discussion on effective temperature, see the Refs .~\cite {Hoelck:2016tqf, Nendzig:2014qka, Singh:2018wdt, Singh:2021evv, Singh:2025xrd}. 

\item $m_{\rm D}$ is the Debye mass. In the deconfined medium, the Debye mass has contributions from both the gluon and quark degrees of freedom, and is expressed as;
\begin{equation}
m_{\rm D}(T) = T_{\rm eff} \sqrt{4\pi\alpha_{s}(\Lambda^{2}) \left( \frac{N_c}{3} +\frac{N_f}{6} \right)}
\label{mDeB}
\end{equation}
here, $\alpha_{s}(\Lambda^{2})$  is the temperature-dependent strong running coupling constant, 
\begin{equation}
\alpha_{s}(\Lambda^{2})  = \frac{1}{b_1\ln\left(\frac{\Lambda^{2}}{\Lambda_{\overline{MS}}^2}\right)} 
\label{alpha}
\end{equation}

where $b_{1}=\frac{11N_c - 2N_f}{12\pi}$, $\Lambda_{\overline{MS}}$ = 0.176 GeV for  $N_f = 3$ and renormalization scale $\Lambda$  = $2 \pi T$.

\item $\alpha_{\rm eff}$ is effective coupling constant, depending on the strong
coupling constant at soft scale $\alpha_{s}^{s} =
\alpha_{s} (m_{c} \alpha_{s}/2)$, given as  $\alpha_{\rm eff} = \frac{4}{3}\alpha_{s}^{s}$.
\end{itemize}

The medium temperature evolution or medium cooling T($\tau$, $b$) is obtained by solving the second-order relativistic viscous hydrodynamics equations under the consideration of a vortical medium~\cite{Sahoo:2023xnu}. In Ref.~\cite{Sahoo:2023xnu}, the medium cooling law is investigated by incorporating the effect of vorticity and viscosity on the hydrodynamic evolution of hot QCD matter. The present work adopts the same hydrodynamic formulation to obtain the temperature evolution in the presence of vorticity $\omega$, which is given by:

\begin{align}
\label{tempevolution}
    \frac{dT}{d\tau} = \frac{1}{\gamma}\left[- \frac{T}{3}\bigg( 1 + \frac{2\omega T^{2}}{s\pi^{2}}\rm \cosh\left(\frac{\omega}{2T}\right) \bigg) +  \frac{\Phi T^{-3}} {12 \mathit a} \right]\partial_{\mu} u^{\mu}
\end{align}
\noindent
where, $ u^{\mu} =\gamma(1, \vec{v})$ is the four-velocity vector, with $\gamma = \frac{1}{\sqrt{1-\vec{v}^{2}}}$ being the Lorentz factor. Here, $s = c+dT^{3}$ is the entropy density obtained by fitting the quasiparticle model equation of state, the $c$ and $d$ being fit parameters with values 0.04829 GeV$^{3}$ and 16.46, respectively. The constant, $a$ in Eq.~\ref{tempevolution}, is defined as;

\begin{equation}
  a = \frac{\pi^{2}}{90} \left[ 16 + \frac{21}{2}N_{f} \right] \nonumber
 \end{equation}
where $N_{f} = 3$, is the number of flavors.\\

In Eq.~\ref{tempevolution}, $\Phi = \pi^{00} - \pi^{zz}$ is the difference between temporal and spatial components of the shear viscosity tensor $\pi^{\mu \nu}$ representing the viscous term. The evolution of the shear viscous term in the presence of vorticity is given by;

\begin{align}
 \frac{d\Phi}{d\tau} &= - \frac{2aT\Phi}{3b\gamma} - \frac{\Phi}{2\gamma}\left( \partial_{\mu}u^{\mu} - \frac{5\gamma}{T}\frac{dT}{d\tau} \right) \nonumber \\ &+\frac{4aT^{4}}{3\tau\gamma}\left(\bigtriangledown^{<0}u^{0>} - \bigtriangledown^{<z}u^{z>}\right) 
 \label{phievolution}
\end{align}

where $b$ is a constant and defined as;
 
\begin{equation}
 b = (1 + 1.70 N_{f})\frac{0.342}{(1 + N_{f}/6)\alpha_{s}^{2}\ln(\alpha_{s}^{-1})} \nonumber
\end{equation}

The initial condition for viscous term $\Phi_0 = \frac{1}{3\pi}\frac{s_{0}}{\tau_{0}}$ is determined using the entropy density at $\tau = \tau_{0}$.

 Now, in medium vorticity evolution, coupled with the temperature cooling law and viscosity, is given by,

\begin{align}
\label{omegaevolution}
    \frac{d\omega}{d\tau} = \frac{-\pi^{2}}{2G}\bigg[ \frac{4T}{3\gamma}\bigg( s + \frac{2T^{2}\omega}{\pi^{2}} \rm cosh\left(\frac{\omega}{2T}\right) - \frac{\Phi}{T}\bigg)\partial_{\mu} u^{\mu} \nonumber\\
    + \bigg(s+3dT^{3}+\frac{2F}{\pi^{2}}\bigg)\frac{dT}{d\tau}\bigg]
\end{align}

where, $F= 3T^{2}\omega \cosh\left(\frac{\omega}{2T}\right) - \frac{1}{2} \omega^{2}T \sinh\left(\frac{\omega}{2T}\right)$ and
$G = T^{3} \cosh\left(\frac{\omega}{2T}\right) + \frac{1}{2} \omega T^{2}  \sinh\left(\frac{\omega}{2T}\right)$. The initial condition for $\omega$ has been extracted from the chosen values of the circulation parameter $C = 2 \omega \pi r^{2}$ (say r = 1 fm). In addition, the initial condition for vorticity is chosen in such a way that the speed of rotation does not violate causality. The detail derivation of Eqs.(\ref{tempevolution}),~(\ref{phievolution}), and~(\ref{omegaevolution}) can be found in Ref.~\cite{Sahoo:2023xnu}. The initial temperature ($T_0 (b)$) and thermalization time 
($\tau_0$) for QGP is estimated $T_0 (b)$ using a phenomenological relation that connects it to the experimentally measured charged-particle multiplicity as a function of impact parameter, $b$~\cite{Hwa:1985xg}:

\begin{equation}
 T_{0}(b) = \left[\frac{90}{g_{k} 4\pi^{2}} C^{\prime} \frac{1}{A_{T} \tau_{0}} \frac{dN_{\text{ch}}}{dy} \right]^{1/3},
 \label{t0}
 \end{equation}

here the constant is $C^{\prime} = \frac{2\pi^{4}}{45\zeta(3)} \approx 3.6$, and $\frac{dN_{\text{ch}}}{dy} \simeq 
\frac{dN_{\text{ch}}}{d\eta}$, under the massless limit. 
The transverse overlap area, $A_T = \pi R_T^2$, is estimated using the MC-Glauber mode for Pb+Pb collision at 
$\sqrt{s_{\rm NN}} = 5.02$ TeV, where $R_T$ is the transverse radius of the fireball~\cite{Loizides:2017ack}. It is important to emphasize that the initial thermalization time $\tau_0$ plays a decisive role in determining the initial temperature $T_0(b)$. In the present work, we have considered $\tau_0 = 0.2~\text{fm}$ for Pb+Pb collisions at $\sqrt{s_{\text{NN}}} = 5.02~\text{TeV}$; the chosen value is consistent with those used in modern viscous hydrodynamic simulations. Corresponding to this choice of $\tau_0$, the values of $T_0(b)$ are obtained as a function of the measured charged-particle multiplicity. The charged particle multiplicity for Pb+Pb collisions at $\sqrt{s_{\rm NN}} = 5.02$ TeV is taken from ALICE experimental data~\cite{ALICE:2015juo}.\\ 

\begin{figure}
\includegraphics[height = 2.6in, width = 3.0in]{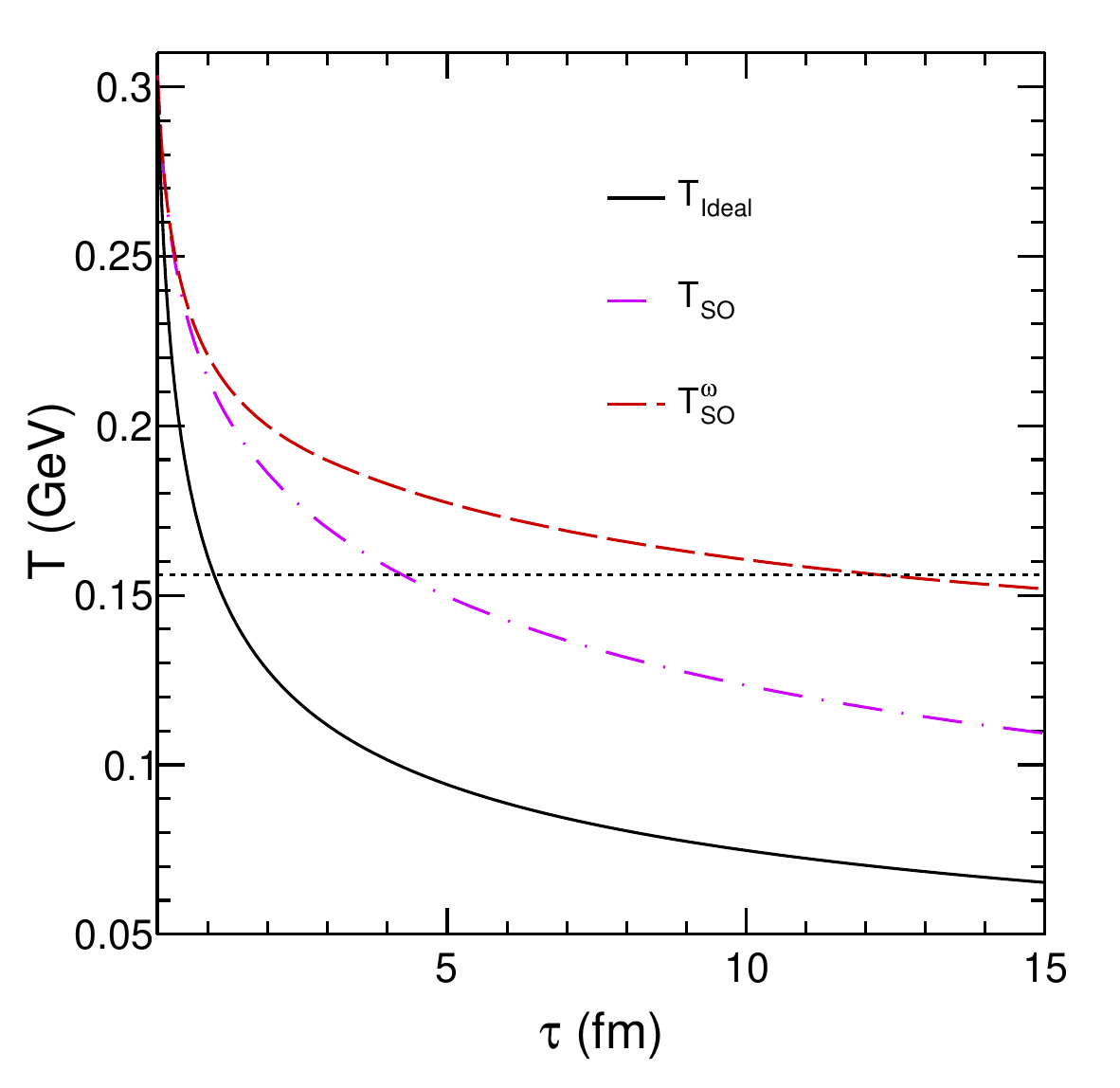}
\caption{(Color online) Evolution of medium temperature $T$ for ideal (solid black line), viscous (dotted dashed magenta), and rotating viscous fluid (dashed red line) with the initial conditions:  T$_{0}$ = 0.300 GeV, 
$\tau_{0}$ = 0.2 fm, $\omega_{0}$ = 0.1 fm$^{-1}$, $\Phi_{0}$ = 0.0516 GeV$^4$.  For T$_{\text{Ideal}}$; $\omega = 0$ and $\Phi = 0$. For T$_{\text{SO}}$; $\omega = 0$ but $\Phi \ne 0$. For T$^{\omega}_{SO}$; $\omega \ne 0$ and $\Phi \ne 0$. The black dashed line at T = 0.156 GeV shows the chemical freeze-out hypersurface. }
\label{fig:0}
\end{figure}

\begin{figure}
\includegraphics[scale=0.4]{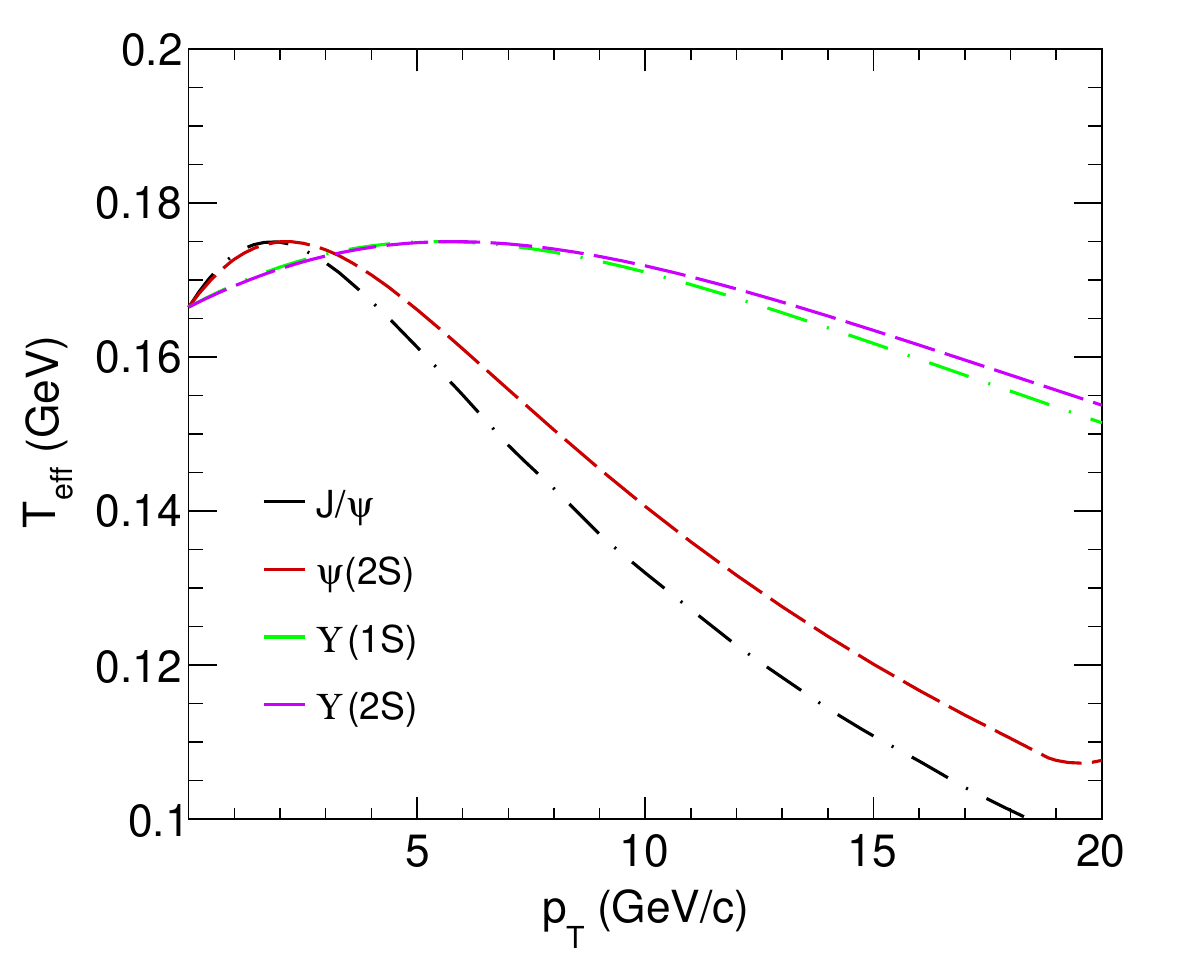}
\caption{(Color online) The effective temperature ($T_{\rm eff}$) as a function of $p_{\rm T}$ for $J/\psi$, $\psi$(2S), $\Upsilon$(1S), and $\Upsilon$(2S) states calculated at medium temperature $T$ = 0.175 GeV.}
\label{fig:0b}
\end{figure}

In Fig.~\ref{fig:0}, the temperature cooling law as a function of proper time $\tau$ is shown for certain cases; for the ideal case ($T_{\rm Ideal}$: non-viscous and non-vortical fluid), the second-order viscous correction to the cooling law ($T_{\rm SO}$ finite viscosity but fluid is non-vortical) and for the case where fluid is vortical and have finite viscosity ($T_{\rm SO}^{\rm \omega}$). A rapid cooling is observed for ideal fluid in the absence of any dissipative effects, as shown in Fig.~\ref{fig:0}.  However, in the presence of viscosity, additional heat is brought to the system, which further slows down the temperature cooling. Moreover, for the case of viscous-vortical fluid, the viscosity coupled with vorticity significantly enhances the QGP lifetime compared to non-vortical viscous and non-viscous cases. Further, the obtained medium temperature is used to estimate the effective temperature. The variation of $T_{\rm eff}$ as a function of $p_{T}$ for different quarkonium states is depicted in Fig.~\ref{fig:0b}. Fig.~\ref{fig:0b} is obtained at medium temperature T = 0.175 GeV. The initial rise in the $T_{\rm eff}$ at $p_{T}$ around 1 to 3 GeV for charmonium and 1 to 6 GeV for bottomonium states is the consequence of the blue-shift that arises at $v_{Q} \ll v_m$. At $v_{Q} \gg v_m$ or high-$p_{T}$ red-shift dominates and leads to a decrease in the $T_{\rm eff}$ for both charmonium and bottomonium states. The change in the pattern of the $T_{\rm eff}$ shown in Fig.~\ref{fig:0} for $J/\psi$, $\psi$(2S), $\Upsilon$(1S), and $\Upsilon$(2S) is a reflection of the difference in their masses, which is explicitly discussed in Ref.~\cite{Captain:2025jtv}.

\begin{figure*}
\includegraphics[scale=0.4]{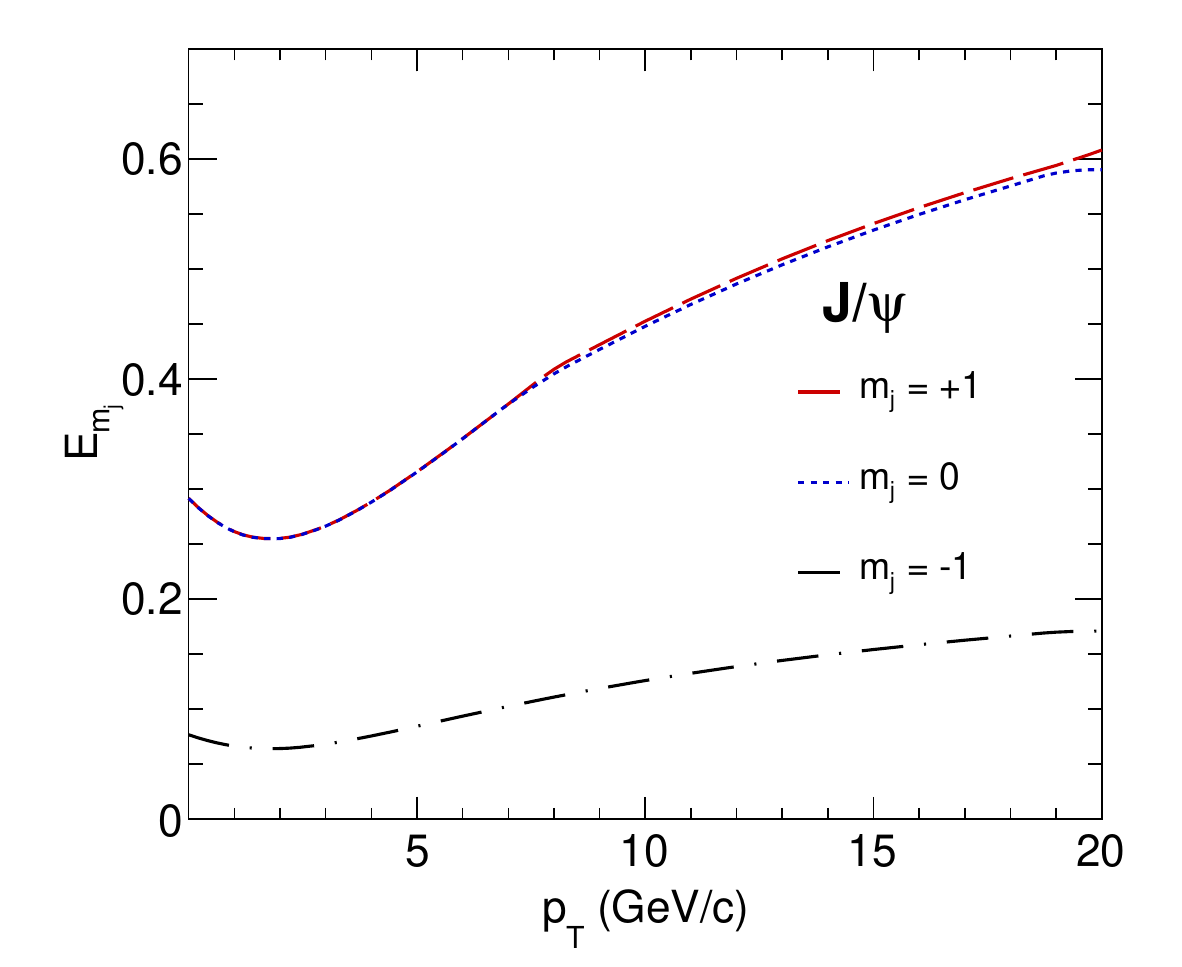}
\includegraphics[scale=0.4]{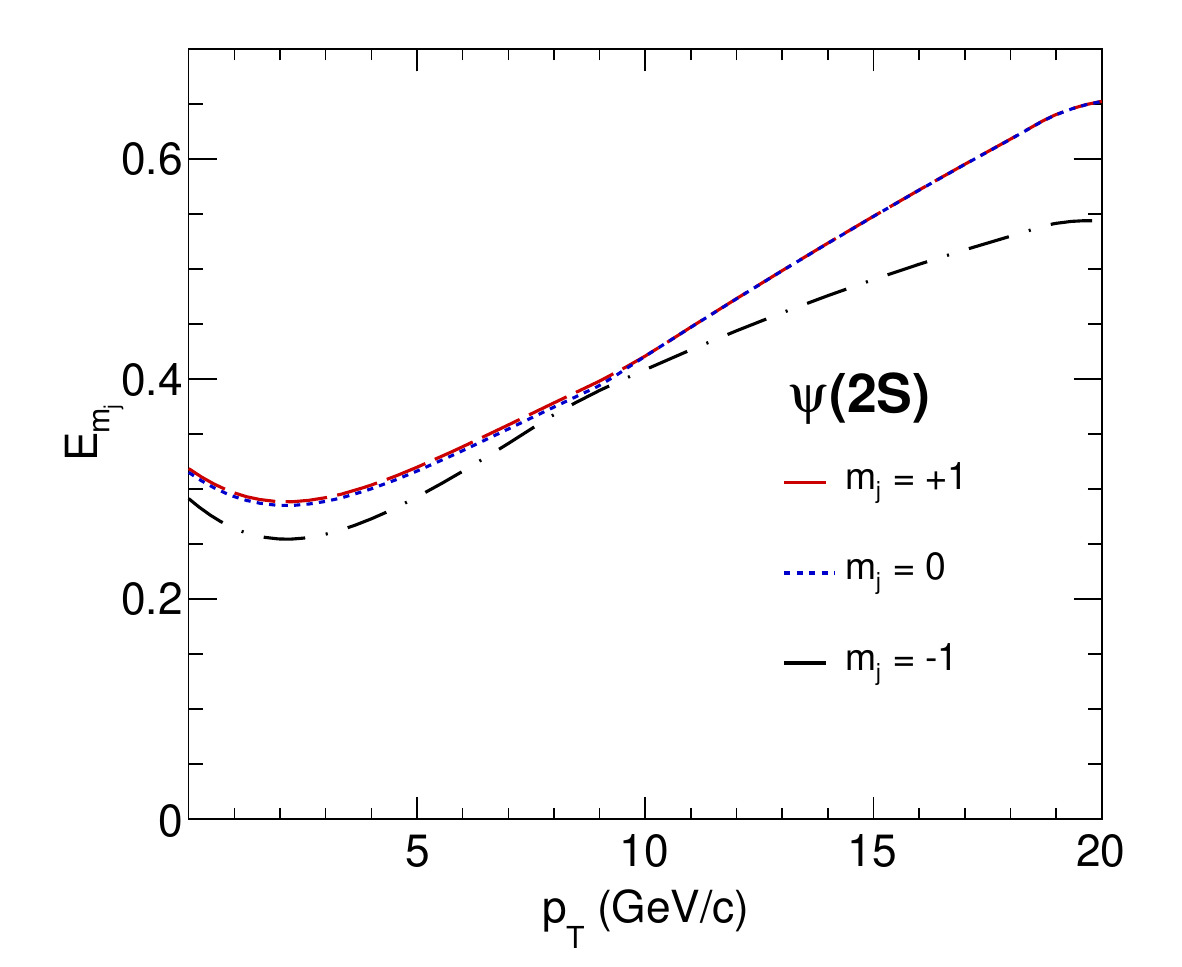}
\includegraphics[scale=0.4]{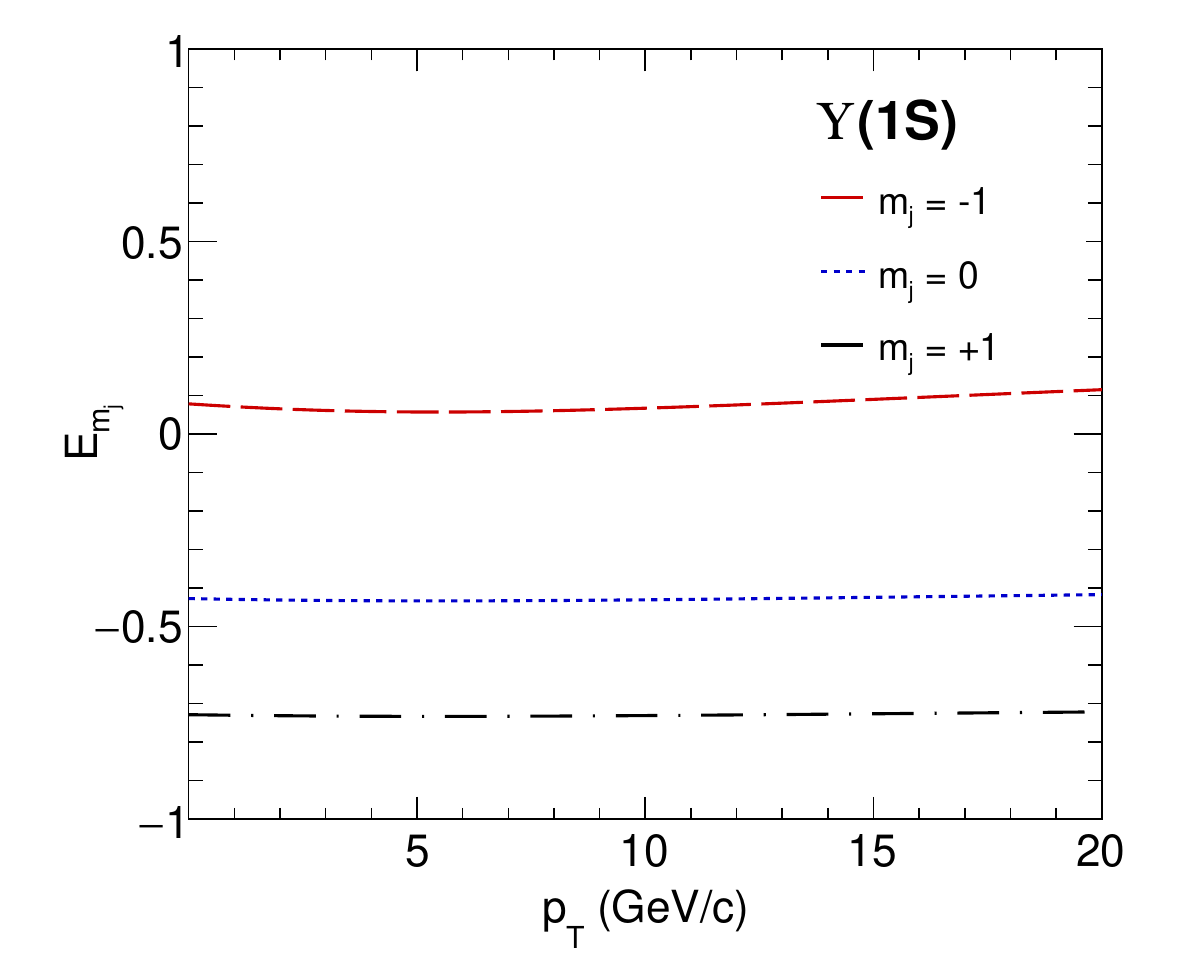}
\includegraphics[scale=0.4]{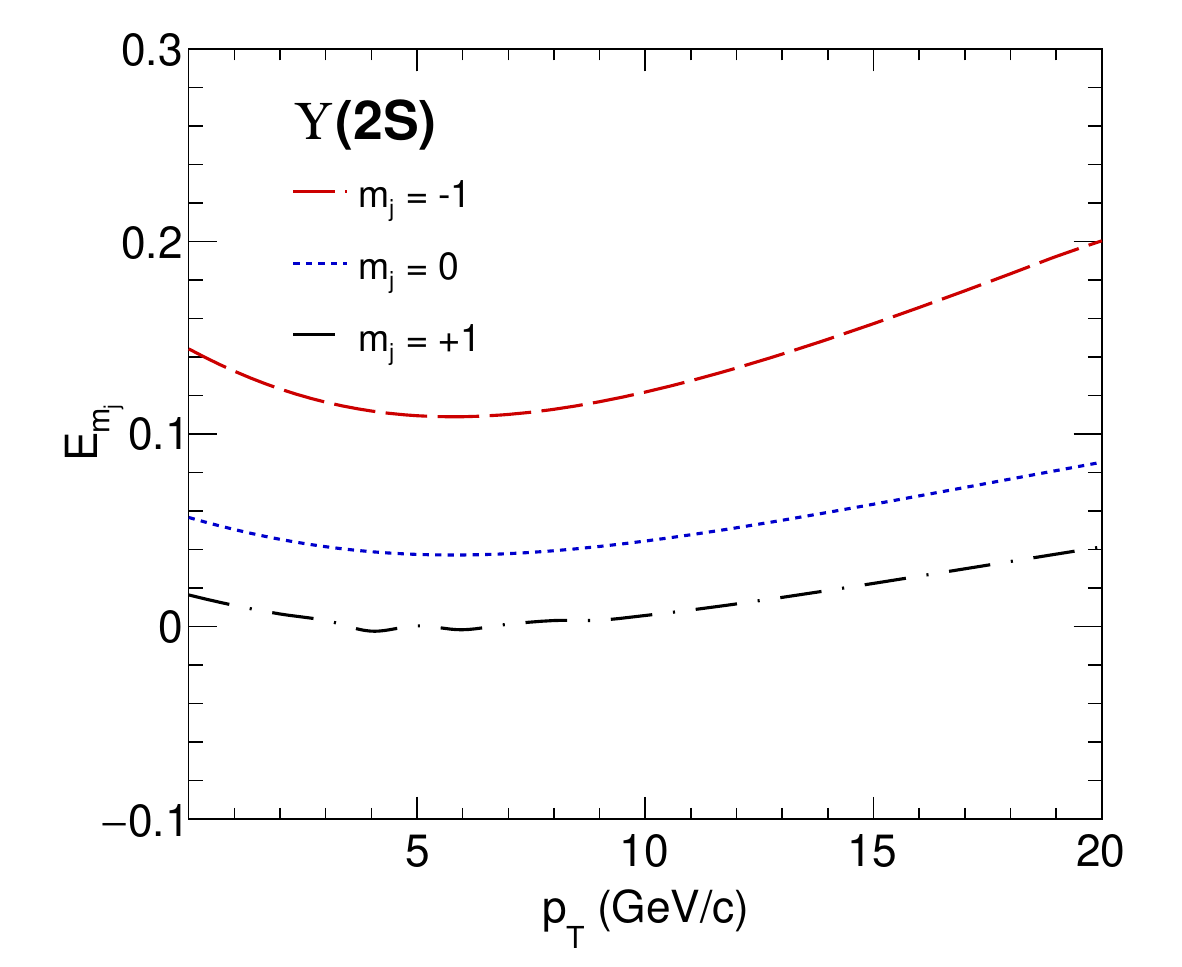}
\caption{(Color online) The  energy eigenvalue as a function of $p_{\rm T}$  for $J/\psi$,  and $\psi$(2S) states (upper panel), $\Upsilon$(1S), and $\Upsilon$(2S) states (lower panel) obtained at $C$ = 1.0 fm using $T$ = 0.175 GeV.}
\label{fig:01}
\end{figure*}

We solve Eq.~\ref{shrr2} numerically by taking a $10^4$ point logarithmically spaced finite spatial grid. The quarkonium wave functions and corresponding energy eigenvalues are obtained using a temperature-dependent real part ($\Re[V(r, m_{\rm D})]$) of the complex quarkonium potential as defined in Eq.~\ref{potential}. Figure \ref{fig:01} shows the energy eigenvalues for $J/\psi$, $\psi (2S)$, $\Upsilon (1S)$, and $\Upsilon (2S)$ states are obtained as a function of $p_{\rm T}$ for three different spin projection quantum number $m_{j}$ values at C = 1.0 fm and  $T$ = 0.175 GeV. Figure \ref{fig:01} indicates that the energy eigenvalue strongly depends on $m_{j}$. Particularly, for $J/\psi$ resonance, for $m_{j}$ = -1, the energy eigenvalue decreases considerably as compared to $m_{j}$ = 0. While for $m_{j}$ = 1, the energy eigenvalue slightly increases with respect to $m_{j}$ = 0 at higher $p_{\rm T}$. However, for $\Upsilon (1S)$ and $\Upsilon (2S)$, a clear splitting of the energy eigenvalues is observed for three different spin projections because of the spin-vorticity coupling. We found that the rotation in the medium leads to an increase (decrease) of the eigenenergy depending on the value of $m_{j}$ = 1 ($m_{j}$ = -1) as compared to the irrotational case, i.e., $m_{j}$ = 0. The upper and lower panel of Fig.~\ref{fig:01} shows that the medium modification alters eigenenergy differently for different quarkonium states, and thus it is natural to expect that the degree of spin alignment will be different for different quarkonium states.\\

The relativistic corrections to the potential model are neglected during the evolution of quarkonium states due to the large mass of charm and bottom quarks. As Eq.~\ref{potential} indicates, the energy eigenvalues and quarkonium wave function depend on the Debye mass. Hence, in principle, any change in the Debye mass will change the eigenvalues and wave function. In this study, we consider two physical scenarios; first, we discuss the Debye mass in the presence of an external magnetic field, and then the Debye mass in the presence of momentum-space anisotropy.

\subsection{Debye Mass under Strong Magnetic Field}
In the presence of a magnetic field, the change in the Debye mass arises from the quark loop, as gluons do not interact with the magnetic field. The total Debye mass of partons in a strong magnetic field has been calculated using the temporal component of the vacuum self-energy diagram at the static limit. It is given as~\cite{Karmakar:2018aig, Zhang:2020efz}.

\begin{equation}
m_{\rm D}^{2}(T, eB) = 4\pi\alpha_{s}(\Lambda^{2}, |eB|)\bigg[T_{\rm  eff}^{2}\frac{N_c}{3} + \sum_{f} \frac{|q_f eB|}{4\pi^{2}}\bigg].
\label{mDeB}
\end{equation}
Here, $\alpha_{s}(\Lambda^{2}, |eB|)$  is the modified running coupling constant, depending on the magnetic field and temperature. It can be written as~\cite{Ayala:2018wux, Bandyopadhyay:2017cle},

\begin{equation}
\alpha_{s}(\Lambda^{2}, |eB|) = \frac{g_{s}^{2}}{4\pi} = \frac{\alpha_{s}(\Lambda^{2})}{1+b_{1}\alpha_{s}(\Lambda^{2})\ln\left(\frac{\Lambda^{2}}{\Lambda^{2} + |eB|}\right)}
\label{alphaeB}
\end{equation}

\subsection{Effect of  Anisotropy on Debye Mass}
\label{anistropy}
Inhomogeneity in the momentum-space distribution between longitudinal and transverse directions gives rise to the momentum-space anisotropy in relativistic heavy-ion collisions. This anisotropy certainly affects the Debye mass; therefore, in the presence of momentum space anisotropy, the Debye mass is factorized as~\cite{Romatschke:2003ms, Romatschke:2004jh, Strickland:2011aa, Dong:2021gnb, Dong:2022mbo, Strickland:lectures};

\begin{equation}
m_{\rm D}(T,\xi) =  m_{\rm D}(T) \bigg[  f_{1}(\xi) \cos^2\theta + f_{2}(\xi) \bigg]^{-\frac{1}{4}}
\label{mDzeta}
\end{equation}
Here, parameter $\xi$ measures the degree of momentum-space anisotropy, defined as;
\begin{equation}
\xi = \frac{1}{2} \frac{<{\bf{k}_{\perp}^2}>}{<k^2_{z}>} -1
\label{zeta}
\end{equation}
where $k_{z}\equiv\bf{k} \cdot \bf{n}, \bf{k}_{\perp}\equiv \bf{k} - \bf{n}(\bf{k} \cdot \bf{n}$) correspond to the particle momenta along and perpendicular to the direction of anisotropy, and $\theta$ is the angle between particle momentum ($\bf{k}$) with respect to direction of anisotropy ($\bf{n}$). 
Here, $f_{1}(\xi)$ and $f_{2}(\xi)$ are parameterized in such a way that Eq.~\ref{mDzeta} remains always true under small and large values of $\xi$. Here, it is worth mentioning that the parametrization of the anisotropic Debye mass for arbitrary $\xi$ is obtained by constructing a model for the real part of the short-range heavy quark-antiquark potential~\cite{Strickland:2011aa}. 

\begin{align}
f_{1}(\xi) = &\frac{9 \xi (1+\xi)^{\frac{3}{2}}}{2\sqrt{3+\xi}(3+\xi^2)}\nonumber \\ &\times \frac{\pi^2 (\sqrt{2}-(1+\xi)^{\frac{1}{8}} )+ 16 (\sqrt{3+\xi}-\sqrt{2})}{(\sqrt{6}-\sqrt{3})\pi^2 - 16(\sqrt{6}-3)}
\label{f1zeta}
\end{align}

\begin{align}
f_{2}(\xi) =\; \xi \bigg(&\frac{16}{\pi^2}- \frac{\sqrt{2}(\frac{16}{\pi^2}-1)+(1+\xi)^{\frac{1}{8}}}{\sqrt{3+\xi}} \bigg)\bigg( 1 -  \frac{(1+\xi)^{\frac{3}{2}} }{1 + \frac{\xi^2}{3}}\bigg) \nonumber \\ +&   f_{1}(\xi) +1
\label{f2zeta}
\end{align}

It is necessary to mention that in the presence of momentum space anisotropy, the potential becomes non-central; thus solving Eq.~\ref{shrr1} using the separation of variables method will not be applicable.
To retain spherical symmetry of the potential,  we considered a scenario in which the direction of anisotropy is along the direction of the particle,  i.e., $\theta$ = 0. With this assumption, we restore the conditions necessary for solving the Schrödinger equation (Eq.~\ref{shrr1}) in its radial form; here, the medium anisotropy parameter ($\xi$) serves as an arbitrary constant. This approximation effectively captures the leading-order influence of anisotropy while preserving computational feasibility.\\

So far, we have set up the effective Hamiltonian for quarkonium states in a rotating, magnetized, and 
anisotropic QCD medium, incorporating spin-vorticity and spin-magnetic couplings along with the potential. 
It has also been discussed how the Debye mass modifies in the presence of the magnetic field and momentum-
space anisotropy. With this configuration in place, we now discuss the spin alignment of quarkonia in the QGP medium.
\subsection{Quarkonia Spin Alignment}
The spin density operator can be written as
\begin{equation}
\hat{\rho} =\frac{e^{-\beta \hat{H}}}{Z}  ,\; \; \beta = \frac{1}{T}.
\label{rho}
\end{equation}
Here, $\hat{H}$ is the Hamiltonian operator, and Z is the normalization factor corresponding to the spin density matrix, and Z = Tr($\rho$). In general, Eq.~\ref{rho} is the canonical density operator (or density matrix) employed for describing a statistical thermodynamic equilibrium system. Equation~\ref{rho} assumes energy is the only conserved thermodynamic quantity in the system. Additionally, Eq.~\ref{rho} offers several physical interpretations. In the high-temperature limit ($\beta \rightarrow 0$), the canonical ensemble approaches a completely random ensemble where all eigenstates are equally populated. Conversely, in the low-temperature limit ($\beta \rightarrow \infty$), Eq.~\ref{rho} indicates that the canonical ensemble reduces to a pure ensemble, with only the ground state being populated~\cite{Sakurai}.

Recent spin alignment measurements of $J/\psi$ in Pb+Pb collisions at $\sqrt{s_{\rm NN}} = 5.02$ TeV reveal anisotropy in the angular distribution of eigenstates~\cite{ALICE:2020iev, ALICE:2022dyy}. One possible explanation for this anisotropy is the generation of substantial vorticity within the system. Motivated by this, we aim to investigate the anisotropic coefficients in a thermal rotating system, where the form of Eq.~\ref{rho} is modified, and total angular momentum emerges as an additional conserved quantity. \\

The spin density matrix in a rotating frame with angles $\theta_r$ and $\phi_r$ can be transformed as;

\begin{equation}
\hat{\rho}^{r} (\theta_r, \phi_r) = U(\theta_r, \phi_r) \; \hat{\rho} \; U(\theta_r, \phi_r)^{-1} 
\label{rhorot}
\end{equation}

where

\begin{align}
U(\theta_r, \phi_r) = D(\alpha, \beta, \gamma) 
=& \; \exp \left(\frac{-iJ_x \alpha}{\hbar} \right) \exp \left(\frac{-iJ_y \beta}{\hbar} \right)\nonumber \\ & \times \exp \left(\frac{-iJ_z \gamma}{\hbar} \right)
\label{Urot}
\end{align} 
Taking $\alpha = \phi_r, \beta = \theta_r$, and $\gamma = 0$, Eq.~\ref{Urot} can be written as,

\begin{equation}
D^{j}_{m, m^{\prime}}(\phi_r, \theta_r) = \; < j, m | D(\phi_r, \theta_r, 0) | j, m^{\prime} >
\label{Drot}
\end{equation}

with

\begin{equation}
D^{j}_{m, m^{\prime}}(\phi_r, \theta_r) 
= e^{-i m \phi_r} d^{j}_{m, m^{\prime}}(\theta_r) 
\label{Drot2}
\end{equation}

\begin{equation}
\bigg[D^{j}_{m, m^{\prime}}(\phi_r, \theta_r) \bigg]^{-1} 
=  e^{i m \phi_r} \bigg[ d^{j}_{m, m^{\prime}}(\theta_r) \bigg]^{-1} 
\label{Drot3}
\end{equation}

For j = 1, we have explicitly

\begin{equation}
d^{1} (\theta_r) = 
\left( {\begin{array}{ccc}
\frac{1}{2}(1 + \cos \theta_r) &  -\frac{1}{\sqrt{2}} \sin \theta_r &  \frac{1}{2}(1 - \cos \theta_r)   \\
\frac{1}{\sqrt{2}} \sin \theta_r & \cos \theta_r & -\frac{1}{\sqrt{2}} \sin \theta_r \\
\frac{1}{2}(1 - \cos \theta_r) & \frac{1}{\sqrt{2}} \sin \theta_r & \frac{1}{2}(1 + \cos \theta_r)  \\
\end{array} } \right)
\label{adeq3}
\end{equation}

We can expand the right-hand side of Eq.~\ref{rhorot},

\begin{align}
\hat{\rho}^{r}_{m,m^{\prime}} (\theta_r, \phi_r) =  \sum_{m^{\prime \prime}, m^{\prime \prime \prime}} & < j, m | D| j, m^{\prime \prime} > < j, m^{\prime \prime} | \hat{\rho} | j, m^{\prime \prime \prime} >\nonumber \\ &\times < j, m^{\prime \prime \prime} | D^{-1}| j, m^{\prime} > 
\end{align} 

\begin{align}
=\sum_{m^{\prime \prime}, m^{\prime \prime \prime}} e^{-i m \phi_r} d^{j}_{m, m^{\prime \prime}}(\theta_r)& < j, m^{\prime \prime} | \hat{\rho} | j, m^{\prime \prime \prime}> \nonumber \\ &\times e^{i m^{\prime \prime \prime}\phi_r} \bigg[ d^{j}_{m^{\prime \prime \prime}, m^{\prime}}(\theta_r) \bigg]^{-1}
\nonumber
\end{align}

\begin{equation}
= \sum_{m^{\prime \prime}, m^{\prime \prime \prime}} e^{i( m^{\prime \prime \prime} - m)\phi_r} d^{j}_{m, m^{\prime \prime}}(\theta_r) \rho_{m^{\prime \prime}, m^{\prime \prime \prime}} \bigg[ d^{j}_{m^{\prime \prime \prime}, m^{\prime}}(\theta_r) \bigg]^{-1}
\label{finalrho}
\end{equation}
and
\begin{equation}
\rho_{m^{\prime \prime}, m^{\prime \prime \prime}} =   < j, m^{\prime \prime} | \hat{\rho} | j, m^{\prime \prime \prime}> = \frac{1}{Z} e^{-\beta E_{m^{\prime \prime \prime}}} \delta_{m^{\prime \prime},m^{\prime \prime \prime}}
\label{diagonalrho}
\end{equation}

Employing the above-developed formalism, we now estimate the diagonal and off-diagonal elements of the 
spin-density matrix for quarkonium states in a rotating, magnetized, and anisotropic medium. 
The value of $\theta_{r}$ and $\phi_{r}$ used in Eq.~\ref{finalrho} are taken from Ref.~\cite{DeMoura:2023jzz}, where these parameters are determined by solving a system of equations in Collins-Soper frame to $J/\psi$ and $\Upsilon$ states with respective ranges $4.0 < p_{T} < 6.0$ GeV/c and $p_{\rm T} > 15$ GeV/c. The value of $\theta_{r} = 1.75 \pm 0.10$ rad for $J/\psi$ and $\theta_{r} = 1.58 \pm 0.10$ rad for $\Upsilon$ states. The off-diagonal elements are obtained using  $\phi_{r}$ = 0.  We use the freeze-out temperature $T$ = 156 MeV in the calculation of the spin-density matrix elements. This value corresponds to the chemical freeze-out temperature extracted from statistical hadronization model fits to hadron yields measured in heavy-ion collisions at LHC energies~\cite{Andronic:2017pug}. Notably, this temperature is consistent with lattice QCD predictions for the QCD crossover temperature~\cite{Steinbrecher:2018phh}.\\

\section{Results and Discussion}
\label{res} 

\begin{figure}
\includegraphics[scale=0.4]{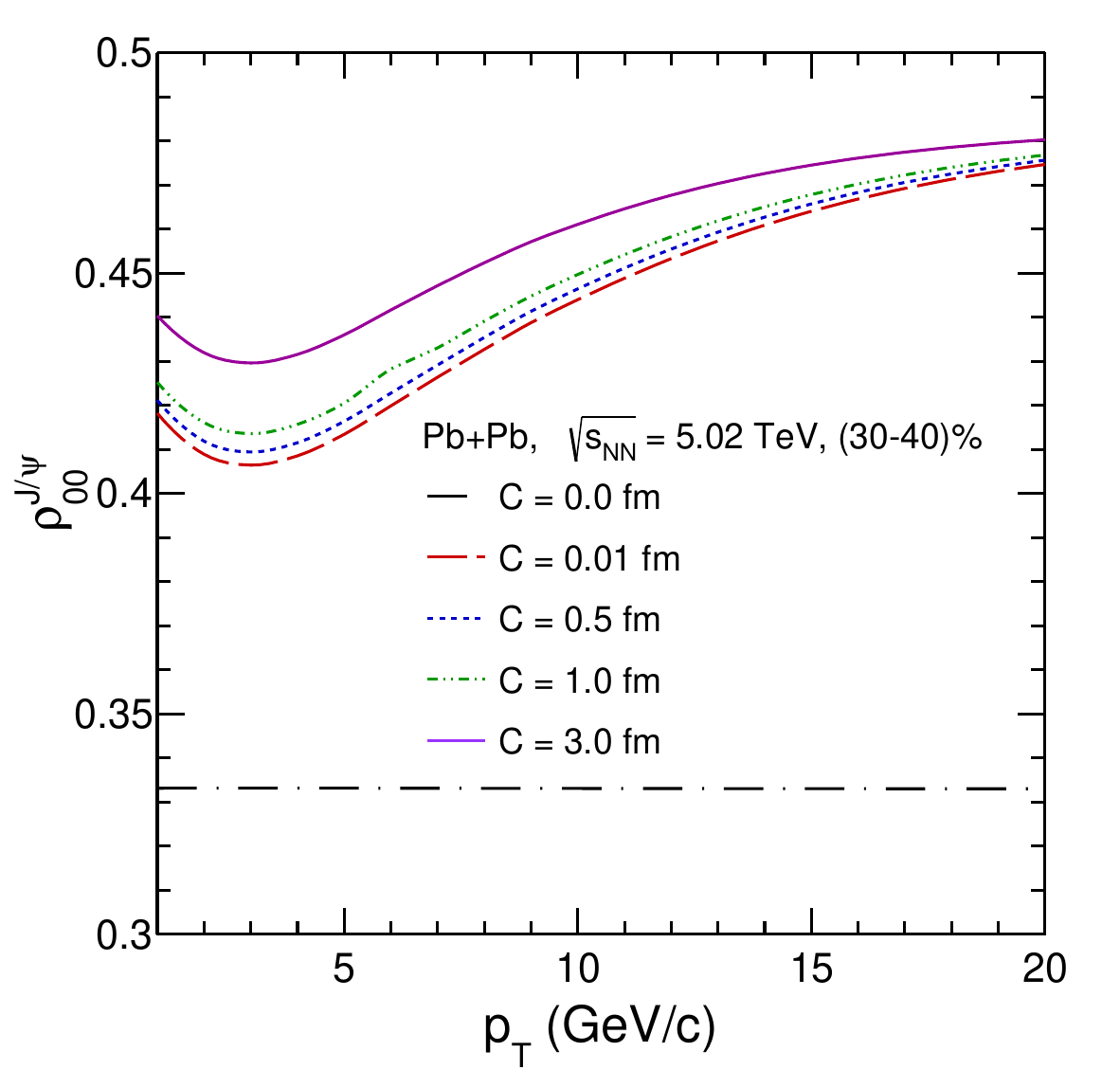}
\includegraphics[scale=0.4]{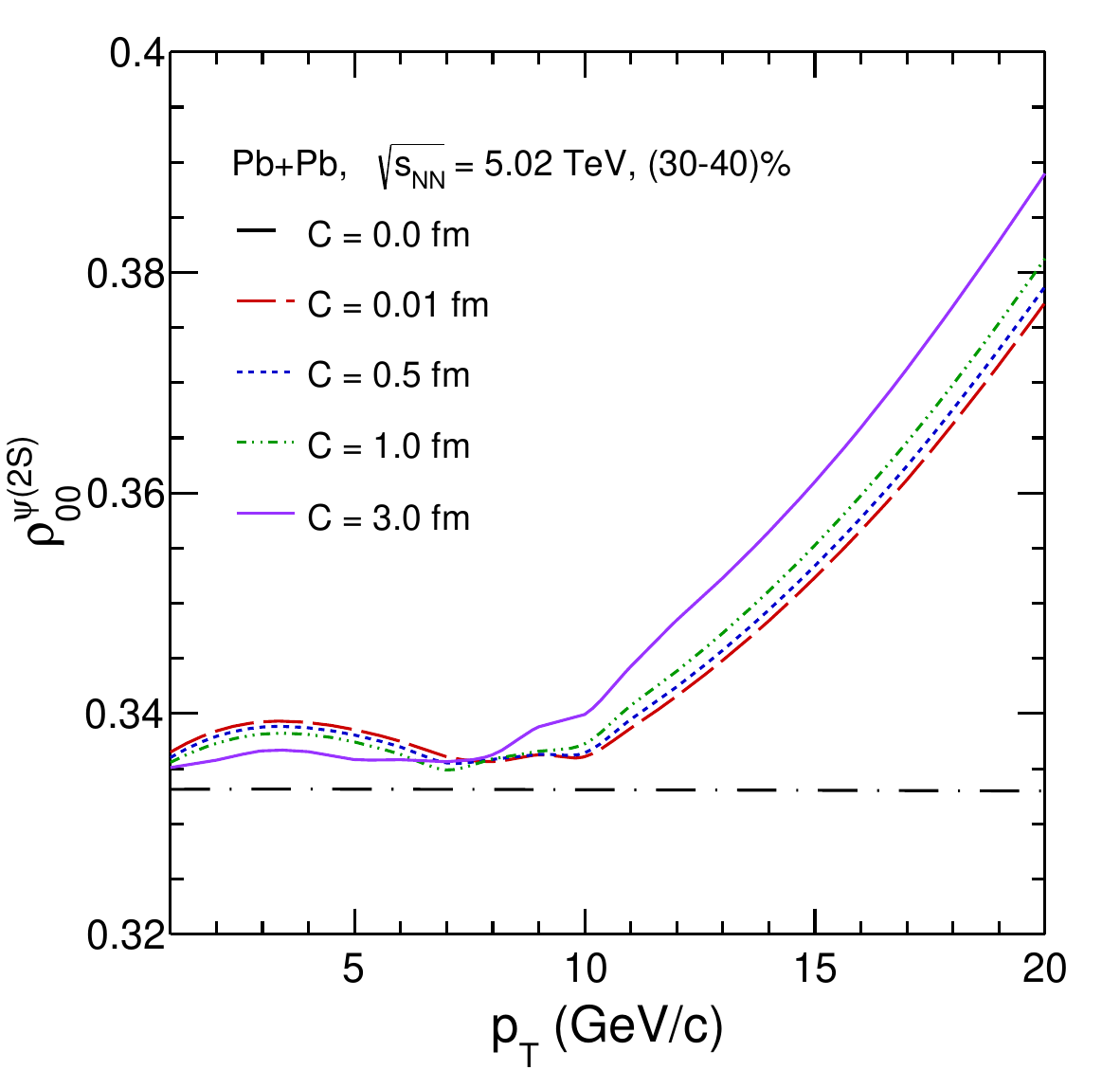}
\caption{(Color online) The spin alignment observable $\rho_{00}$ as a function of $p_{\rm T}$ for $J/\psi$ (upper panel) and $\psi$(2S) (lower panel) in Pb+Pb collisions at $\sqrt{s_{\rm NN}}$ = 5.02 TeV calculated at mid-rapidity for different values of the circulation parameter, $C$.}
\label{fig:1}
\end{figure}

As discussed in the previous section, the energy eigenvalues used to calculate the spin alignment factor depend on the medium temperature, rotation, magnetic field, and anisotropy parameter. The different values of medium temperature, rotation, magnetic field, and anisotropy parameters are considered for this study. 
In this section, first, we will investigate the effect of medium rotation quantified in terms of the circulation parameter $C$ on the diagonal elements of the spin density matrix, i.e., $\rho_{00}$ for different quarkonium states. Then we will investigate the effect of medium temperature ($T$), external magnetic field ($eB$), and momentum-space anisotropy ($\xi$) on $\rho_{00}$ observable. Furthermore, we will explore a few off-diagonal elements of the spin-density matrix for quarkonium states to have a complete understanding of spin-vorticity coupling dynamics in relativistic heavy-ion collisions.\\

\subsection{Diagonal elements}
 Using Eq.~\ref{finalrho}, one can obtain the $00^{\rm th}$ diagonal elements of the spin$-$density matrix in a rotating frame, $\rho^{r}_{00}$, and is expressed as follows;

\begin{equation}
\rho^{r}_{00} = \frac{1}{Z} \bigg[ \cos^{2}{\theta_r} e^{-\beta E_{0}} + \frac{1}{2} \sin^{2}{\theta_r} \left(e^{-\beta E_{1}} + e^{-\beta E_{-1}}\right)\bigg]
\label{diagonalrho}
\end{equation}

\begin{figure}
\includegraphics[scale=0.4]{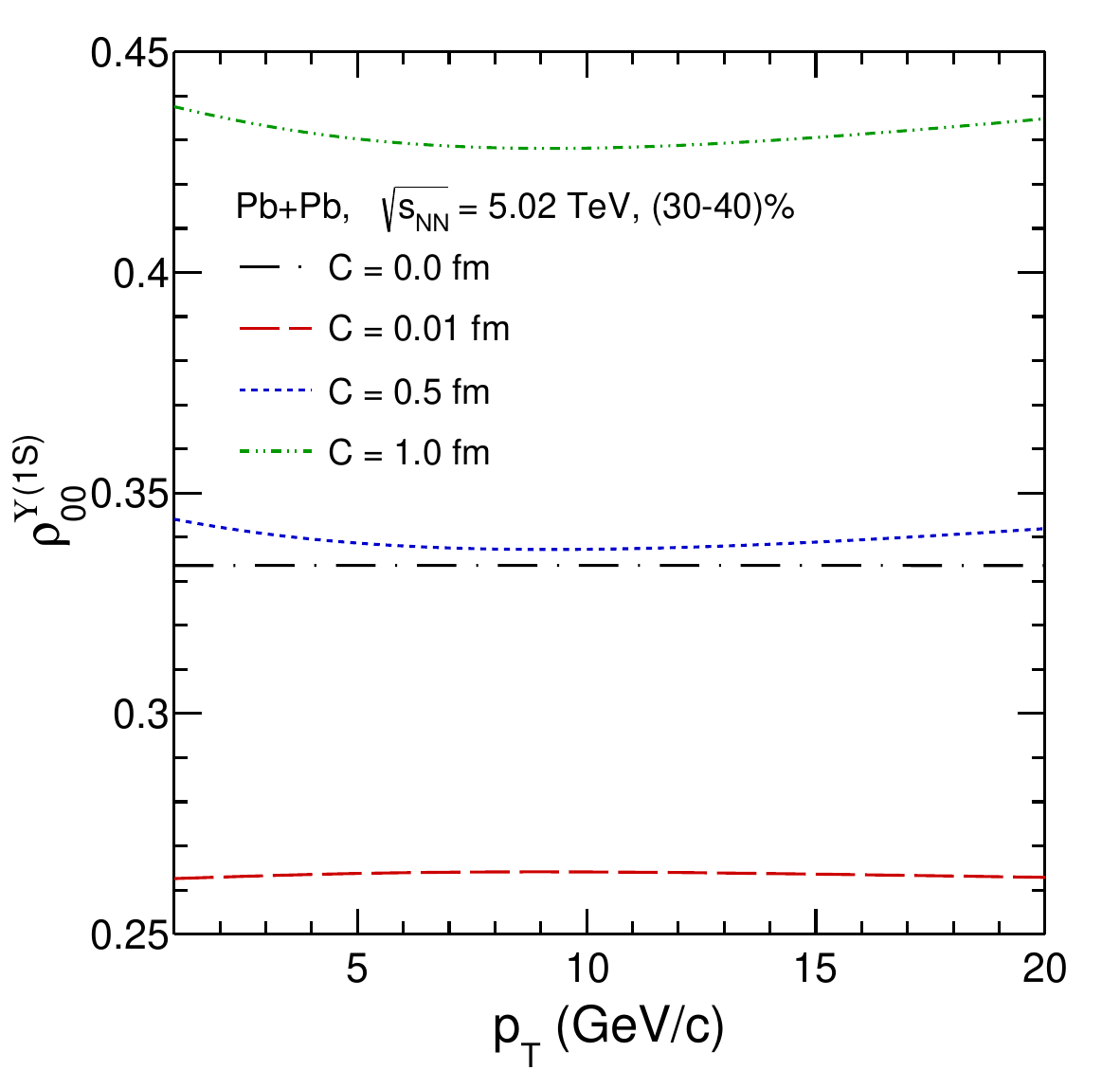}
\includegraphics[scale=0.4]{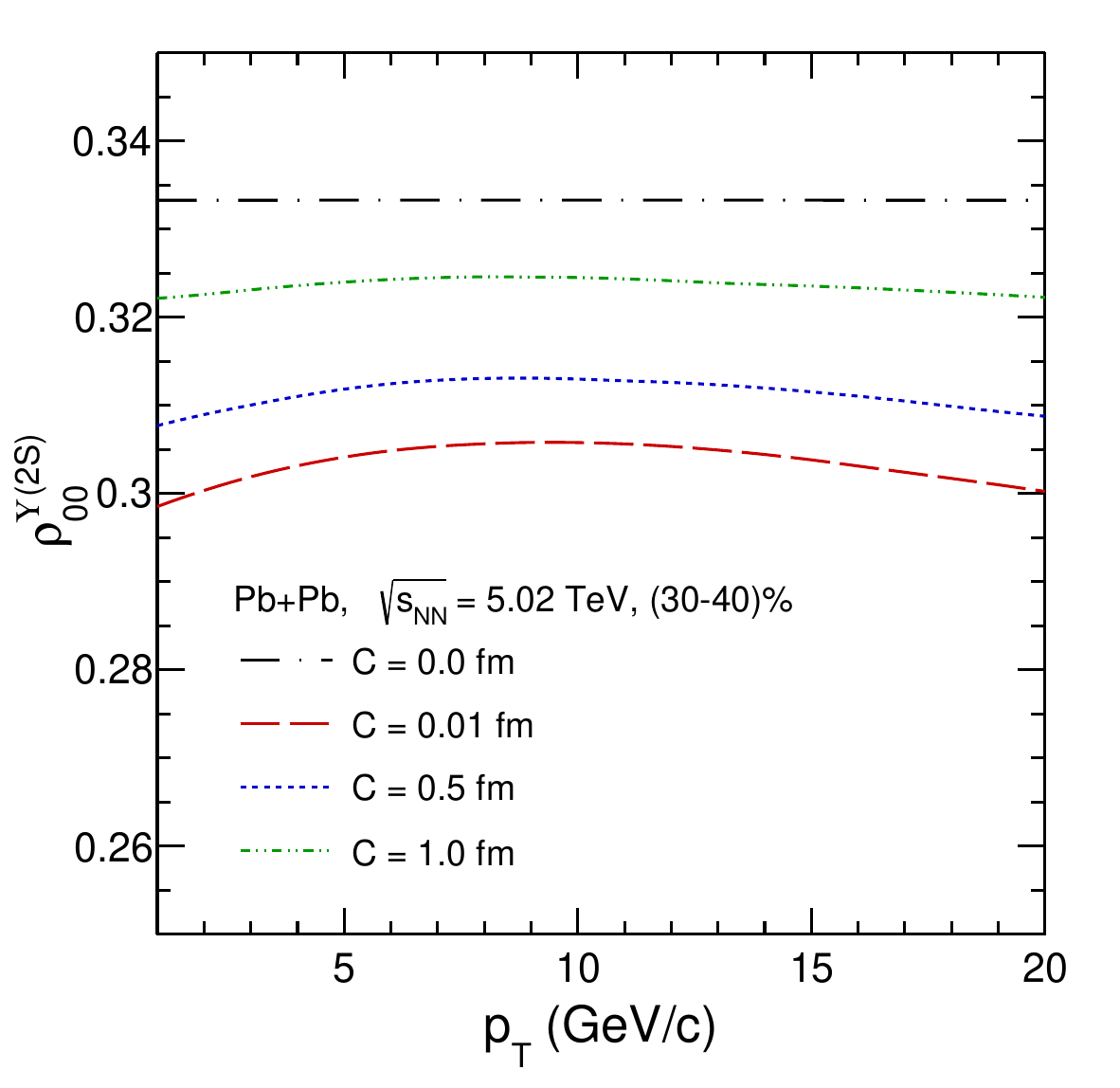}
\caption{(Color online) The spin alignment observable $\rho_{00}$ as a function of $p_{\rm T}$ for $\Upsilon$(1S) (upper), and $\Upsilon$(2S) (lower) in Pb+Pb collisions at $\sqrt{s_{\rm NN}}$ = 5.02 TeV calculated at mid-rapidity for different values of the circulation parameter, $C$.}
\label{fig:2}
\end{figure}

Figure~\ref{fig:1} shows $00^{\rm th}-$elements of the spin$-$density matrix, $\rho_{00}$ \footnote{It is important to note that throughout the study we use  $\rho^{r}_{00}$ as $\rho_{00}$, for simplicity. So, by $\rho_{00}$ we mean the $00^{\rm th}-$ elements of the spin$-$density matrix $\rho$ in rotating frame.},  as a function of transverse 
momentum $p_{\rm T}$ for Pb+Pb collisions at $\sqrt{s_{\rm NN}} = 5.02$ TeV for various values of $C$ at mid-rapidity. The black dashed line in 
Fig.~\ref{fig:1} and other figures as well represent a thermal irrotational medium ($C = 0$), for which $\rho_{00}$ is 1/3. This black dashed line serves as the baseline value for the spin alignment study. Any deviation from the baseline value 
$\rho_{00}$ = 1/3 depicts the degree of alignment of the particle spin in a preferential direction. The upper and lower 
panels of Fig.~\ref{fig:1} demonstrate the $J/\psi$ and $\psi$(2S) spin alignment with $p_{\rm T}$, respectively. Here, the $p_{\rm T}$-dependence in the spin alignment observable $\rho_{00}$ is incorporated through $T_{\rm eff}$, which is used to estimate the quarkonium temperature in the medium. The $T_{\rm eff}$ is a result of relativistic Doppler shift; initially, $T_{\rm eff}$ increases up to $p_{\rm T} \sim 3$ GeV/c and afterward decreases with increasing $p_{\rm T}$ as shown in Fig.~\ref{fig:0}. It has been used to obtain the energy eigenvalue, which is subsequently employed to determine quarkonium spin alignment. As a consequence, the spin alignment observable $\rho_{00}$ for $J/\psi$  initially decreases up to $p_{\rm T} \sim 3$ GeV/c, as shown in the upper panel of Fig.~\ref{fig:1}. At $p_{\rm T} \gtrsim 4$ GeV/c, spin alignment increases with increasing $p_{\rm T}$. Further, for the given $p_{\rm T}$ values, $\rho_{00}$ increases with 
increasing circulation parameter $C$. The information about vorticity is encoded in the conserved circulation parameter, $C$. 
The coupling between quarkonium spin and medium vorticity enhances the quarkonium stability in the medium 
by lowering its eigenenergy. The effect of medium vorticity was found to be marginal for $\omega \leq 0.16$ 
fm$^{-1}$ (corresponds to $C = 1$ fm, and $r =$ 1 fm). However, with increasing vorticity or $C$, $\rho_{00}^{J/\psi}$ increases without changing the pattern as a function of $p_{\rm T}$. \\

The spin alignment of $\psi$(2S) is shown in the lower panel of Fig.~\ref{fig:1}, demonstrates a slight increase in 
$\rho_{00}$ at $p_{\rm T} \leq 2$ GeV/c. At these low $p_{\rm T}$ values, the  $\psi$(2S) is supposedly moving slower 
than the surrounding medium, resulting in a reduced $\rm T_{\rm eff}$ which in turn reduces the eigenenergy, leading 
to an increase in $\rho_{00}$. Consequently, we observe an increasing trend in $\rho_{00}$ with the initial $p_{\rm T}$ 
values. For the intermediate  $ p_{\rm T}$ range, i.e. $2 < p_{\rm T} < 9$ GeV/c,  $\psi$(2S) moves more coherently with 
medium. This results in an increase in $\rm T_{\rm eff}$ as shown in Fig.~\ref{fig:0} and so the eigenenergy, causing $\rho_{00}$ to decrease 
within this $p_{\rm T}$ range. The $\psi$(2S) being an excited state has a relatively high eigenenergy that makes it 
relatively unstable and less susceptible to polarize in the medium. Hence, its spin alignment remains small compared to 
$J/\psi$. In terms of the circulation parameter, $C$,  $\psi$(2S) shows a different spin alignment pattern than 
$J/\psi$, at $p_{\rm T} \lesssim 9$ GeV/c, i.e., the $\rho_{00}$ matrix element decreases with $C$. This can be attributed to a complex interplay between medium vorticity and 
$\rm T_{\rm eff}$. However, for $p_{\rm T} > 9$ GeV/c, the spin alignment behavior of $\psi$(2S) tends to follow a trend 
similar to $J/\psi$, although the relative difference of $\rho_{00}$ from the baseline value is small for $\psi$(2S). \\

 Similarly, in Fig.~\ref{fig:2} we have shown the spin alignment of $\Upsilon$(1S) (upper panel), and $\Upsilon$(2S) (lower panel) for Pb+Pb collisions at $\sqrt{s_{\rm NN}} = 5.02$ TeV  for various values of $C$  at mid-rapidity. As $\Upsilon$(1S) is a massive and tightly bound state, it has small eigen energies, which give rise to the large spin polarization for very small and large values of $C$, i.e., $C$ =  0.01 fm and 1.0 fm. Around $C = 0.5$ fm, it predicts negligible polarization for $\Upsilon$(1S), as $\rho_{00} \approx \frac{1}{3}$. The infinitesimal dependence of $\rho_{00}$ on $p_{\rm T}$ shows that $\Upsilon$(1S) is a highly stable particle, and any changes brought in its eigenenergy by $ T_{\rm eff}$ at $T$ = 0.175 GeV remain almost ineffective. On the other hand, $\rho_{00}$ for $\Upsilon$(2S) shown in the lower panel of Fig.~\ref{fig:2} varies with $p_{\rm T}$. The spin alignment pattern of $\Upsilon$(2S) follows the same explanation as given for $\psi$(2S) at $p_{\rm T} < 9$ GeV/c. However, for $\Upsilon$(2S) state $p_{\rm T}$ range is extended to $p_{\rm T} \geq 20$ GeV/c because of its larger mass compared with $\psi$(2S). The degree of spin alignment is found to decrease with the considered circulation parameter $C$.

It is noteworthy to mention, that  $\rho_{00}$ is more sensitive to $C$ for $\Upsilon$(1S) state compared to $\Upsilon$(2S) states. For example, for $\Upsilon$(1S) state at $C$ = 0.01 fm, the deviation of $\rho_{00}$ from the baseline value 1/3 ($\rho_{00} < 1/3 $) is smaller, while at $C$ = 1.0 fm, it is greater than 1/3 ($\rho_{00} > 1/3 $). It indicates that the shape of the angular distribution of the $\Upsilon$
state flips with the anisotropy created due to the medium vorticity. Since $\rho_{00}$ is closely related to the polarization parameter $\lambda_{\theta}$, this change reflects a transition from transverse to longitudinal spin polarization. The value of vorticity at which this transition occurs is highly sensitive to the temperature and equilibration time of the quark-gluon plasma. Hence, the spin alignment (or angular distribution) of the $\Upsilon$(1S) state may serve as a probe of system thermalization and could potentially be used to estimate the equilibration time. The deviation of $\rho_{00}$ from the 1/3 baseline value is negative for the $\Upsilon$(2S) state for all considered values of $C$. However, there is a clear trend that with increasing $C$, the $\rho_{00}$ value approaches 1/3. At higher $C$, it is expected $\Upsilon$(2S) has similar features as $\Upsilon$(1S).\\

\begin{figure}[ht!]
\includegraphics[scale=0.4]{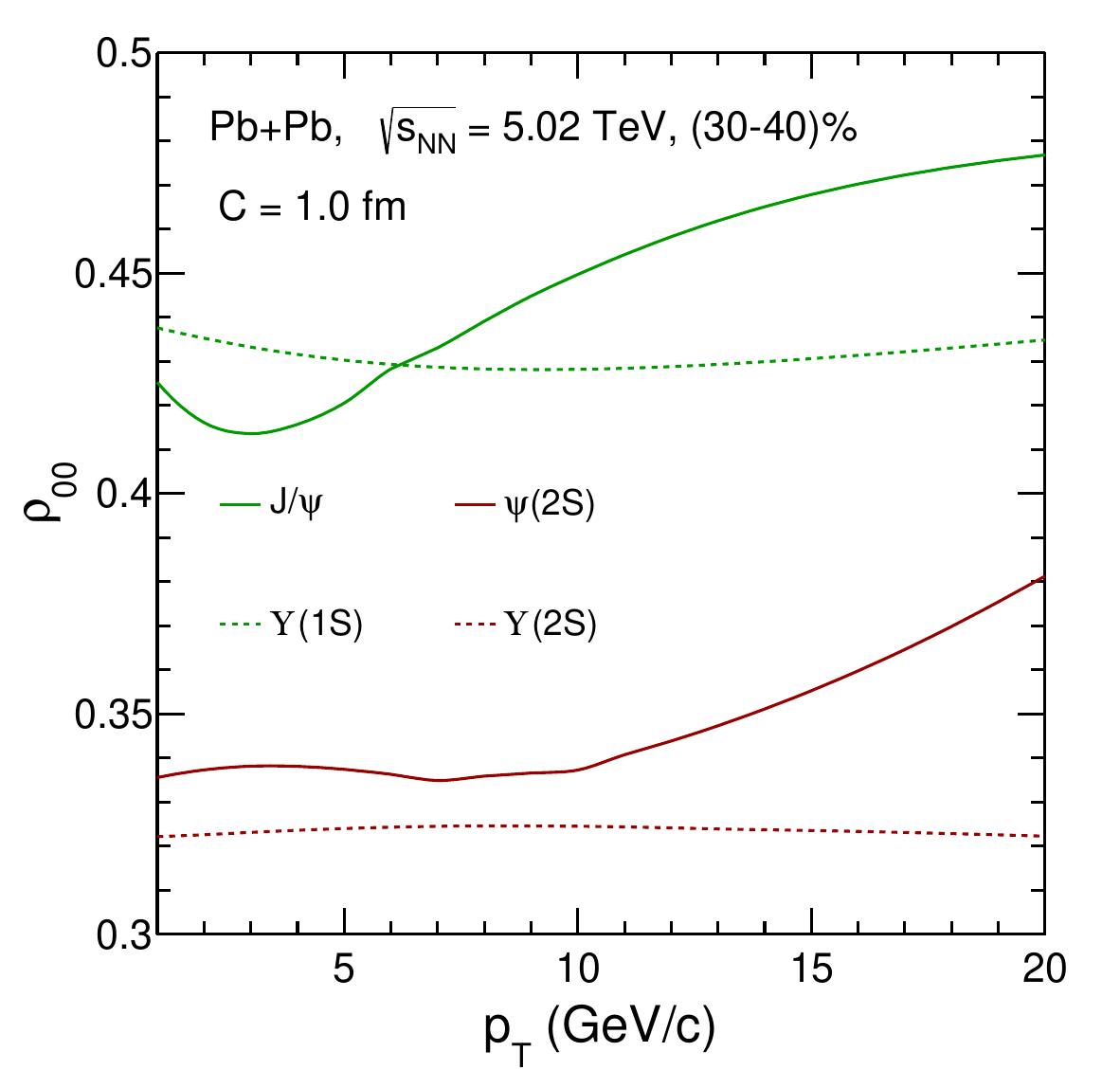}
\caption{(Color online) The spin alignment observable $\rho_{00}$ as a function of $p_{\rm T}$ for $J/\psi$, $\psi$(2S), $\Upsilon$(1S), and $\Upsilon$(2S) states  for Pb+Pb collisions at $\sqrt{s_{\rm NN}}$ = 5.02 TeV calculated at mid-rapidity at the circulation parameter, $C$ = 1.0 fm.}
\label{fig:3}
\end{figure}

A comparison for the order of spin alignment for $J/\psi$, $\psi$(2S), $\Upsilon$(1S), and $\Upsilon$(2S) as the function of $p_{\rm T}$  for Pb+Pb collisions at $\sqrt{s_{\rm NN}}$ = 5.02 TeV calculated at mid-rapidity is shown in Fig.~\ref{fig:3} at $C$ = 1.0 fm. The $p_{\rm T}$ dependence of $\rho_{00}$ observable for different quarkonium states can be understood from the eigenenergy trend as a function of $p_{\rm T}$ (see Fig.~\ref{fig:01}). The interplay of the eigenenergies due to three spin projections and their net $p_{\rm T}$ dependence is reflected in the $\rho_{00}$ observable. In particular, charmonium states exhibit an increasing trend in $\rho_{00}$ with $p_{\rm T}$, indicating that unperturbed particles are most likely to polarize in the medium. While $\rho_{00}$ for bottomonium states remains constant over whole $p_{\rm T}$. At $p_{\rm T} \lesssim 7$ GeV/c, the order of spin alignment of $J/\psi$ is smaller than $\Upsilon$(1S), because perturbation in the state is large at low $p_{\rm T}$, which decreases with increasing  $p_{\rm T}$. As a result, for $p_{\rm T} > 7$ GeV/c, the order of spin alignment of $J/\psi$ becomes larger than $\Upsilon$(1S). It can also be inferred that relatively lighter mass particles with almost the same binding energy are more prone to polarize in the medium. Among all, $J/\psi$ shows the largest deviation from the unpolarized baseline ($\rho_{00} = 1/3$) at high $p_{\rm T}$, followed by $\Upsilon$(1S) state. While $\psi$(2S) shows a moderate spin alignment and $\Upsilon$(2S) remains the least polarized, with $\rho_{00}$ values closer to 1/3.\\ 

\begin{figure}
\includegraphics[scale=0.4]{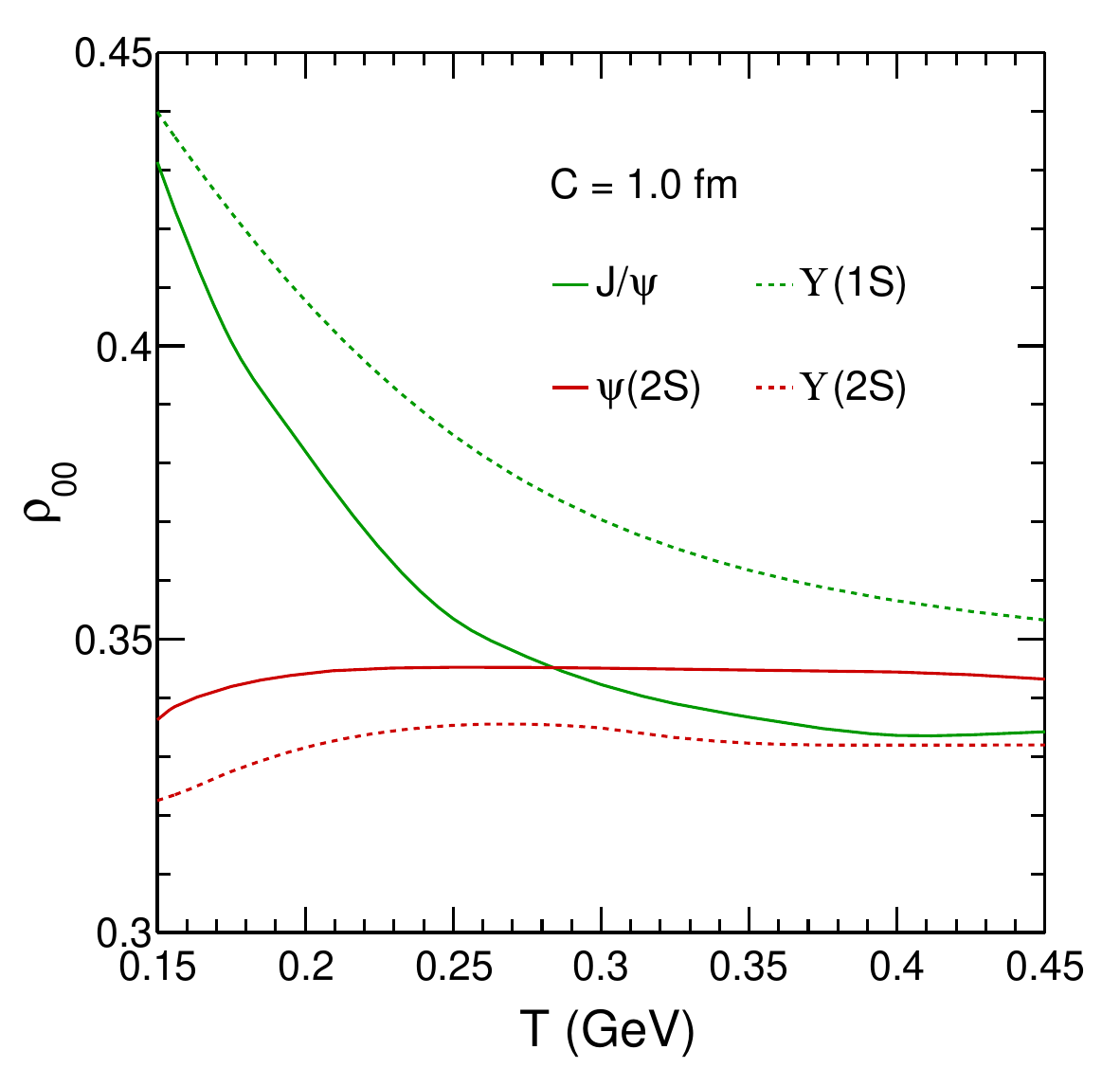}
\includegraphics[scale=0.4]{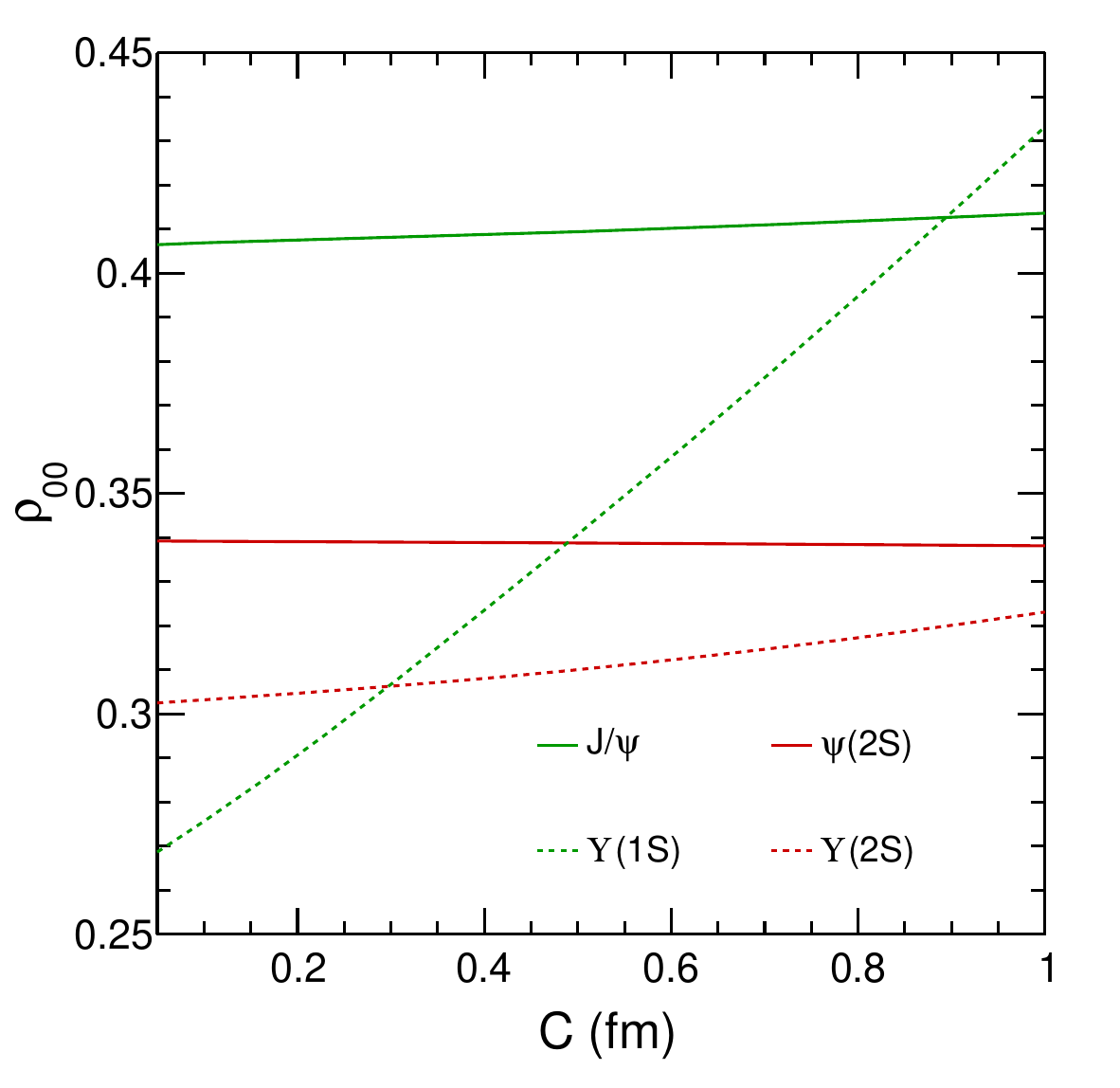}
\caption{(Color online) The spin alignment of quarkonium states such as  $J/\psi$, $\psi$(2S), $\Upsilon$(1S), and $\Upsilon$(2S) states are shown in the upper and lower panels of the figure. The upper panel represents the spin alignment observable $\rho_{00}$ as a function of temperature $T$, which acts as a proxy for charged particle multiplicity,  at a conserved circulation $C$ = 1.0 fm. The lower panel depicts the spin alignment as a function of $C$  for Pb+Pb collisions at $\sqrt{s_{\rm NN}}$ = 5.02 TeV, calculated at mid-rapidity. }
\label{fig:4}
\end{figure}

Figure~\ref{fig:4} depicts the spin alignment when particles are at rest in the medium and explicitly explores the temperature and vorticity dependence. The upper panel of Fig.~\ref{fig:4} displays the variation of $\rho_{00}$ as a function of temperature $T$ for $J/\psi$, $\psi$(2S), $\Upsilon$(1S), and $\Upsilon$(2S) states at $C = 1.0$ fm. The results indicate that the spin alignment of the quarkonium 1S states exhibits a notable sensitivity to the medium temperature. In particular, $\rho_{00}$ decreases with increasing temperature for both $J/\psi$ and $\Upsilon$(1S). This behavior can be attributed to the rapid increase in the magnitude of the eigen energies of these states at high temperatures, which renders them progressively unstable. However, $\Upsilon$(1S), having a higher binding energy compared to $J/\psi$, retains a finite spin alignment even at high temperatures, with $\rho_{00} > \frac{1}{3}$. In contrast, the spin alignment for $J/\psi$ diminishes with temperature, approaching the unpolarized limit of $\rho_{00} \sim \frac{1}{3}$. We found that the deviation of $\rho_{00}$ from 1/3 is a small positive value for $\psi$(2S) and has a mild dependence as a function of temperature. This is because the energy eigenvalues have a marginal dependence as a function of temperature. Similarly, for the $\Upsilon$(2S) states, $\rho_{00}$ exhibits only a marginal dependence on temperature up to $T \approx 270$ MeV, beyond which it tends to saturate towards 1/3.\\

The lower panel of Fig.~\ref{fig:4} illustrates the dependence of $\rho_{00}$ on the parameter $C$ for the aforementioned quarkonium states for Pb+Pb collisions at $\sqrt{s_{\rm NN}}$ = 5.02 TeV calculated at mid-rapidity. For the $J/\psi$ resonance, $\rho_{00}$ increases with $C$, whereas a slight decreasing trend is observed for $\psi$(2S). In the case of $\Upsilon$(1S) and $\Upsilon$(2S), $\rho_{00}$ increases monotonically with increasing $C$. These observations demonstrate that the spin alignment behavior varies across different quarkonium states, despite all being vector mesons. This variation primarily arises due to differences in meson mass and the consequent changes in eigenenergy under the influence of vorticity. As discussed in Fig.~\ref{fig:2}, among all the resonances, Fig.~\ref{fig:4} depicts a clear sensitivity of $\Upsilon$(1S), and $\Upsilon$(2S) state to the circulation parameter $C$. For $C < 0.8$ fm, $J/\psi$ exhibits a higher degree of spin alignment than the other mesons. Conversely, $\Upsilon$(2S) displays the lowest spin alignment at $C = 0.8$ fm, with $\rho_{00} \sim \frac{1}{3}$, which can be ascribed to its relatively large mass and comparable eigen energies for three spin projection states. Moreover, the findings suggest that increasing vorticity has a pronounced effect on the energy eigenstates of heavier particles and therefore reinforces their spin alignment. Consequently, for $C > 0.8$ fm, the spin alignment of $\Upsilon$(1S) and $\Upsilon$(2S) surpasses charmonium counterpart $J/\psi$ and $\psi$(2S), respectively.\\

Although the same observable $\rho_{00}$ is studied in Ref.\cite{DeMoura:2023jzz} as a function of circulation parameter $C$, for $C = 1$, the authors have found
no spin alignment for quarkonia. However, in the present study, at $C = 1$ fm we predict a finite spin alignment. This is potentially due to the medium-modified potential. 

\begin{figure}
\includegraphics[scale=0.4]{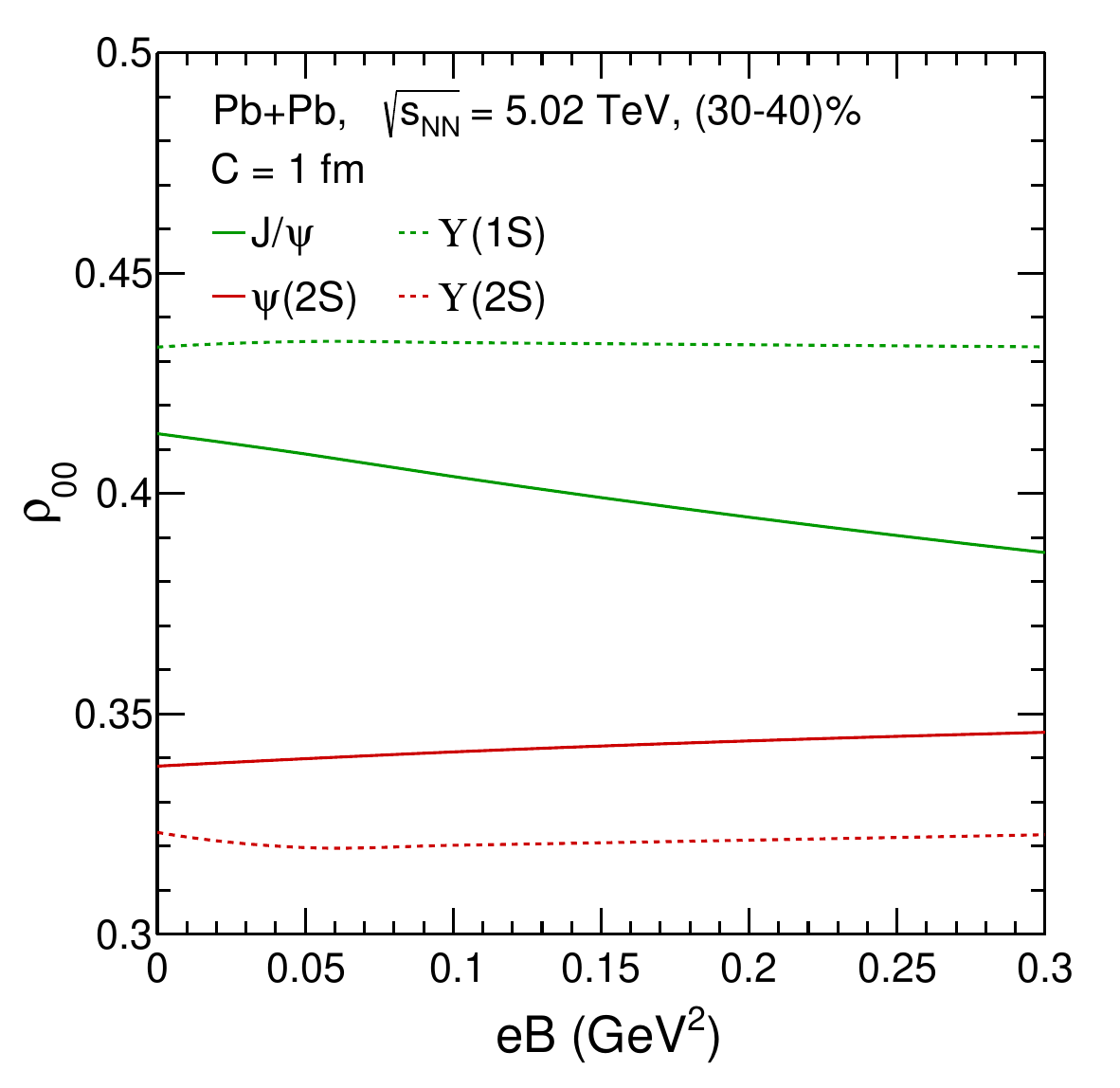}
\includegraphics[scale=0.4]{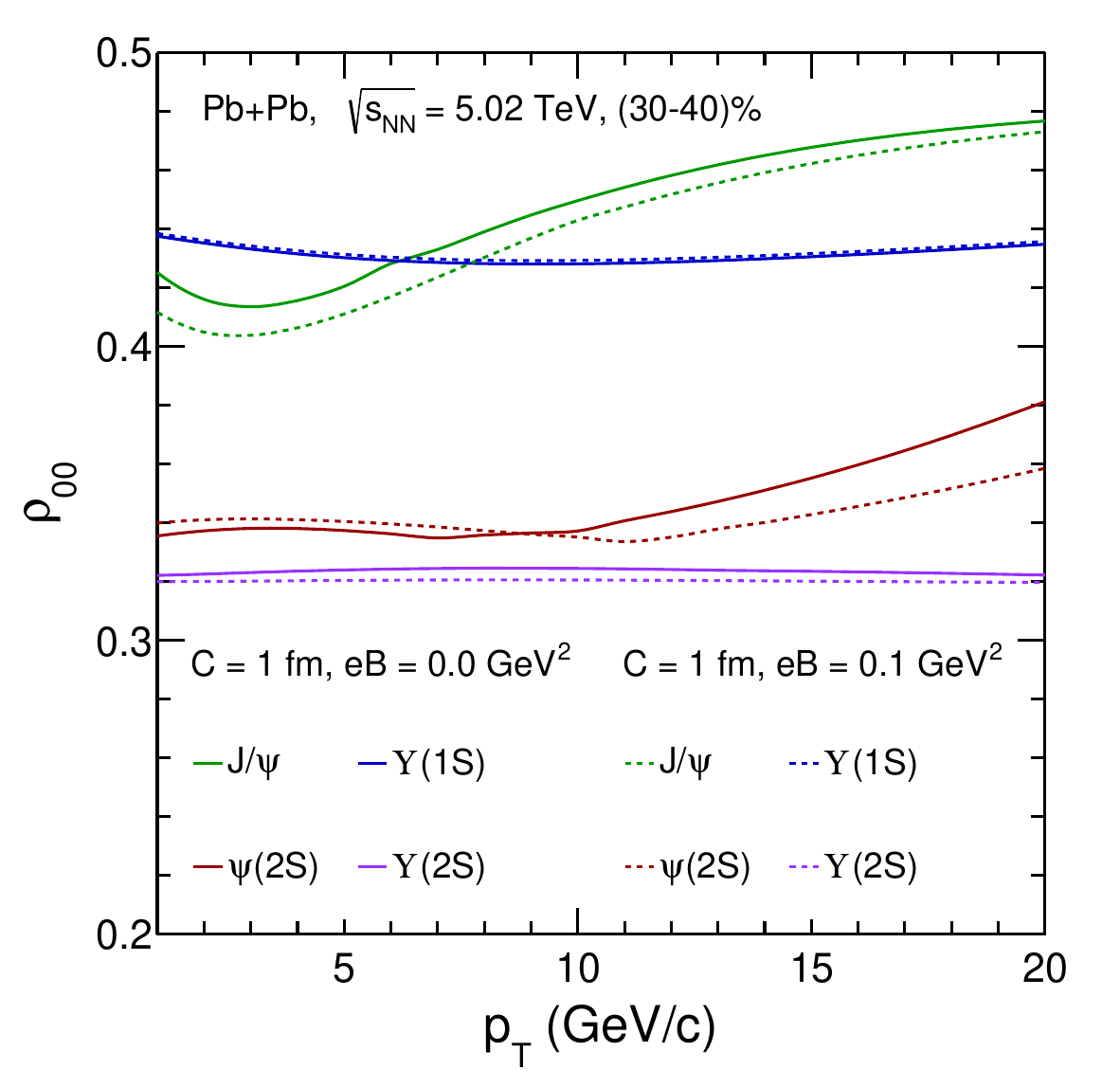}
\caption{(Color online) The spin alignment observable $\rho_{00}$ as a function of $p_{\rm T}$  (upper panel) and magnetic field $eB$ (lower panel) for $J/\psi$, $\psi$(2S), $\Upsilon$(1S), and $\Upsilon$(2S) states  for Pb+Pb collisions at $\sqrt{s_{\rm NN}}$ = 5.02 TeV calculated at mid-rapidity at $C$ = 1.0 fm.}
\label{fig:5}
\end{figure}

Besides the vorticity, a strong magnetic field is considered to be produced in the heavy-ion collisions and is expected to affect the spin alignment of vector mesons~\cite{Sahoo:2024yud}. Since heavy quarks (charm and/or beauty) are primarily produced at the early stage of heavy-ion collisions, the initially generated magnetic field is anticipated to polarize the heavy quarks. Hence, heavy quarks can be considered a crucial probe to study the spin dynamics and magnetic field effect in ultra-relativistic heavy-ion collisions. In Fig.~\ref{fig:5}, we investigate the initial state dynamics, such as the effect of rotation and magnetic field on the spin alignment observables. The interplay of coupling of spin with rotation ($\boldsymbol{\omega} \cdot \mathbf{S}$) and magnetic field ($\boldsymbol{\mu} \cdot \mathbf{B}$, where $\boldsymbol{\mu} = g\frac{q}{2m}\boldsymbol{S}$), is examined through the spin density matrix element $\rho_{00}$.\\ 

The upper panel of Fig.~\ref{fig:5} shows $\rho_{00}$ as a function of $eB$ at $C$ = 1.0 fm in Pb+Pb collisions at $\sqrt{s_{\rm NN}}$ = 5.02 TeV calculated at mid-rapidity for 1S and 2S quarkonium states. Our study predicts a nominal decrease in the order of spin alignment for $J/\psi$ with an increasing magnetic field. While the spin alignment of  $\psi$(2S) almost remains invariant with the magnetic field. 
The magnetic field is found to marginally affect the spin alignment of $\Upsilon$(1S) as the magnitude of $\rho_{00}$ remains almost constant with the magnetic field. Likewise, $\Upsilon$(2S) also seems to be largely unaffected by the magnetic field and left unpolarized, as $\rho_{00}$ for $\Upsilon$(2S) closely varies around 1/3. The change in  $\rho_{00}$ for all the chosen quarkonium states with $eB$ follows a similar explanation as Fig.~\ref{fig:4}, where the eigenenergy was enhanced due to temperature, and here the magnetic field does the same. Unlike Fig.~\ref{fig:4}, the subtle behavior of $\rho_{00}$ can be considered as an artifact of the change brought to the coupling constant, $\alpha_{s}(\Lambda^{2},eB)$, in the presence of the magnetic field.\\

The lower panel of Fig.~\ref{fig:5} illustrates $\rho_{00}$ as a function of $p_{\rm T}$ at $C$ = 1.0 fm for the same collision system and energies. It is shown to quantify the impact of the magnetic field on the spin alignment of various quarkonium states. Therefore, it is shown for two cases; the solid line depicts the presence of a magnetic field $eB$ = 0.1 GeV$^{2}$, whereas the dashed line shows $\rho_{00}$ in the absence of the magnetic field. As the degree of spin-alignment quantified in terms of the deviation of $\rho_{00}$ from 1/3 is decreased for $J/\psi$ due to the magnetic field in the whole $p_{\rm T}$ ranges, while for $\psi$(2S), the deviation of $\rho_{00}$ from  1/3 due to the magnetic field seems to slightly increase at lower $p_{\rm T}$. Interestingly, the magnetic field increases the absolute degree of spin-alignment (i.e., $|\rho_{00} -1/3|$) for $\Upsilon$(1S) and $\Upsilon$(2S) states oppositely. For $\Upsilon$(1S) state $\rho_{00}$ is greater than 1/3, while for $\Upsilon$(2S) state $\rho_{00}$ is less than 1/3. It suggests that, depending on the binding energy of the quarkonium state, the effect of the magnetic field on the spin alignment of the respective state varies accordingly. Here, it is noteworthy to mention the time scale at which quarkonia couple (align) to the vortex (rotation), and the magnetic field of the medium may not necessarily be at the same time. The estimation of the relaxation time for quarkonium spin alignment with the local flow gradients is required for a comprehensive study.\\

\begin{figure}
\includegraphics[scale=0.4]{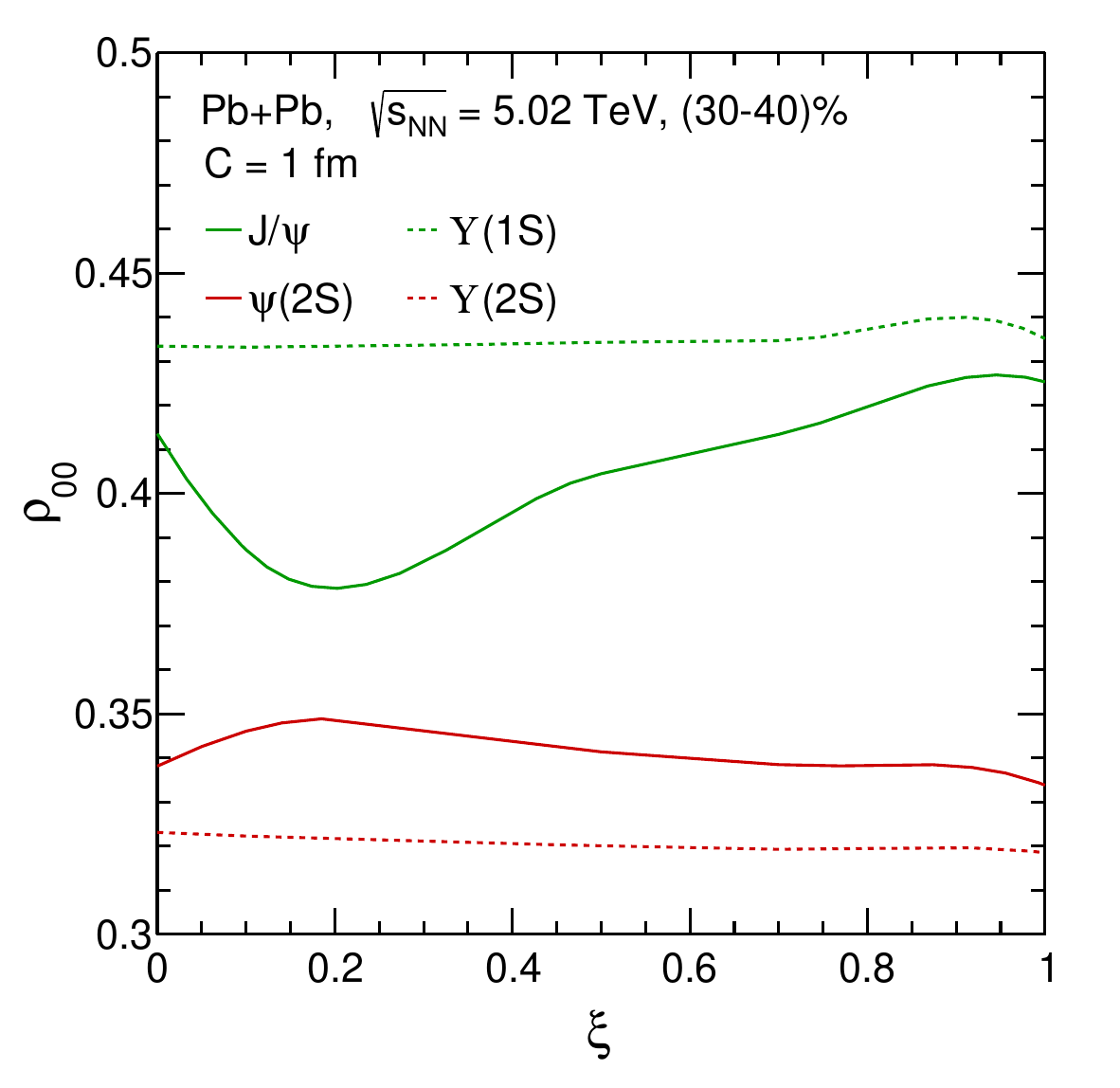}
\includegraphics[scale=0.4]{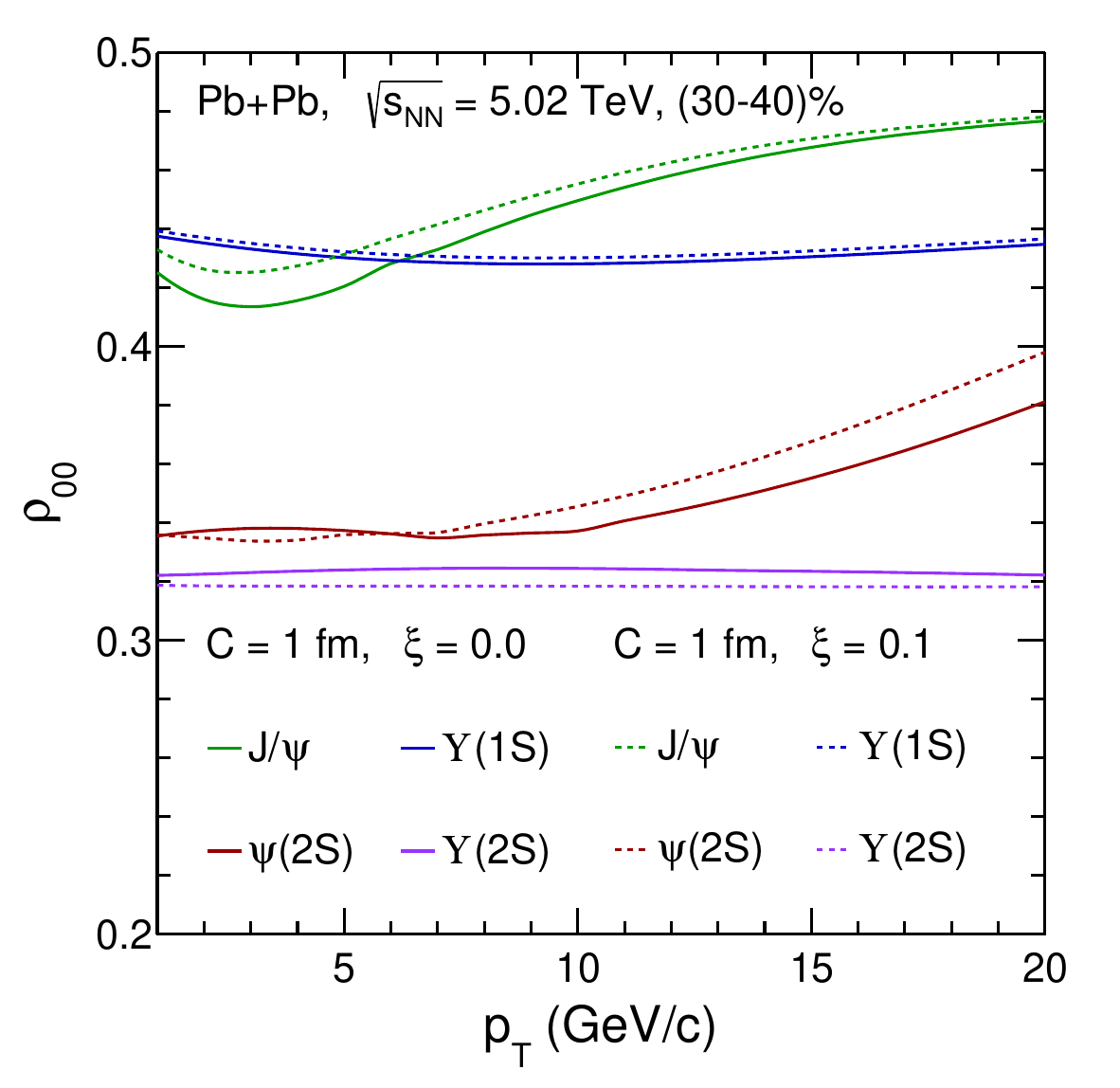}
\caption{(Color online) The spin alignment observable $\rho_{00}$ as a function of $p_{\rm T}$ (upper panel) and medium anisotropy $\xi$ (lower panel) for $J/\psi$, $\psi$(2S), $\Upsilon$(1S), and $\Upsilon$(2S) states  for Pb+Pb collisions at $\sqrt{s_{\rm NN}}$ = 5.02 TeV calculated at mid-rapidity at $C$ = 1.0 fm.}
\label{fig:6}
\end{figure}

As discussed in Section~\ref{anistropy}, anisotropy in the medium is expected to influence the spin alignment of vector mesons~\cite{Sahoo:2024yud}. This effect is explored in Fig.~\ref{fig:6}, which shows how the spin density matrix element $\rho_{00}$ varies with the anisotropy parameter $\xi$. The upper panel of Fig.~\ref{fig:6} presents $\rho_{00}$ as a function of $\xi$ for quarkonium, considering it at rest in the medium, with $C = 1.0$~fm. For small anisotropies ($\xi < 0.2$), a decrease in spin alignment is observed for the $J/\psi$, while the $\psi$(2S) shows only a marginal effect, with a slight enhancement. At higher anisotropy values ($\xi > 0.2$), $\rho_{00}$ for $J/\psi$ increases, whereas for $\psi$(2S), it initially decreases before reaching a saturation point. In contrast, the heavier bottomonium states, $\Upsilon$(1S) and $\Upsilon$(2S), show negligible variation in $\rho_{00}$ across the entire range of $\xi$, indicating that their spin alignment is largely unaffected by the momentum-space anisotropy.\\

The lower panel of Fig.~\ref{fig:6} presents $\rho_{00}$ as a function of $p_{\rm T}$ for various quarkonium states, again at $C = 1.0$~fm. Two scenarios are compared: solid lines represent an anisotropic medium with $\xi = 1.0$, and dashed lines denote the isotropic case ($\xi = 0$). The results reinforce that the impact of anisotropy is state-dependent. For the $J/\psi$, the deviation of $\rho_{00}$ from the isotropic baseline of $1/3$ increases across all $p_{\rm T}$ values in the anisotropic case. Conversely, the $\psi$(2S) shows a slight decrease in deviation at low $p_{\rm T}$. The $\Upsilon$(1S) state remains largely unaffected by either anisotropy or transverse momentum, while for  $\Upsilon$(2S), a small but consistent shift in $\rho_{00}$ is observed in the presence of anisotropy. These findings stress the significance of medium-induced momentum-space anisotropy in modifying the spin alignment of lighter vector mesons. While the effects are modest, they are measurable and offer insights into the early-stage dynamics of the quark-gluon plasma. However, the influence of anisotropy appears minimal for heavier states such as bottomonium, suggesting that their spin alignment serves as a more robust probe that is less sensitive to such medium effects.

\begin{figure*}[ht!]
\centering
\includegraphics[scale=0.29]{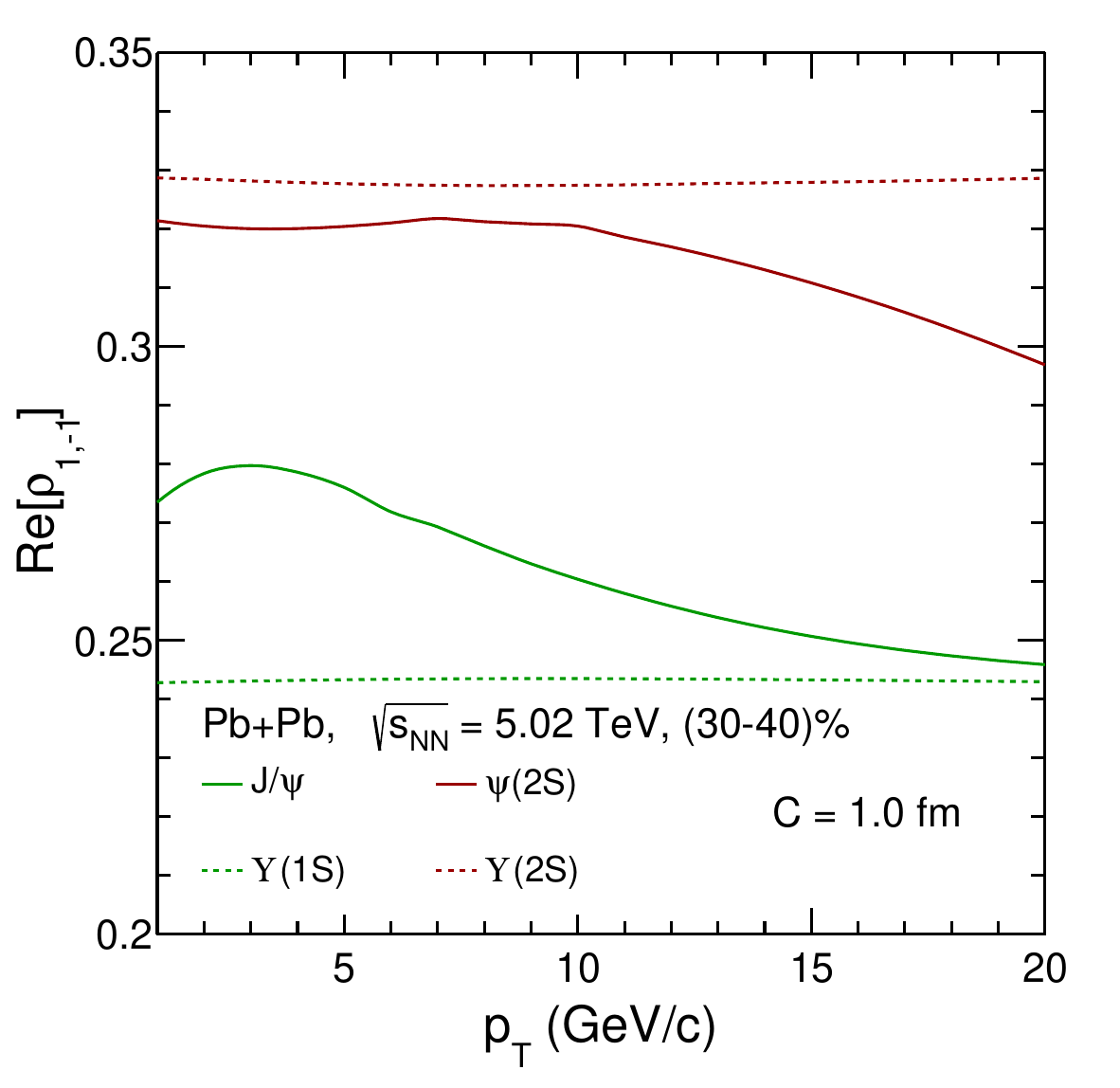}
\includegraphics[scale=0.29]{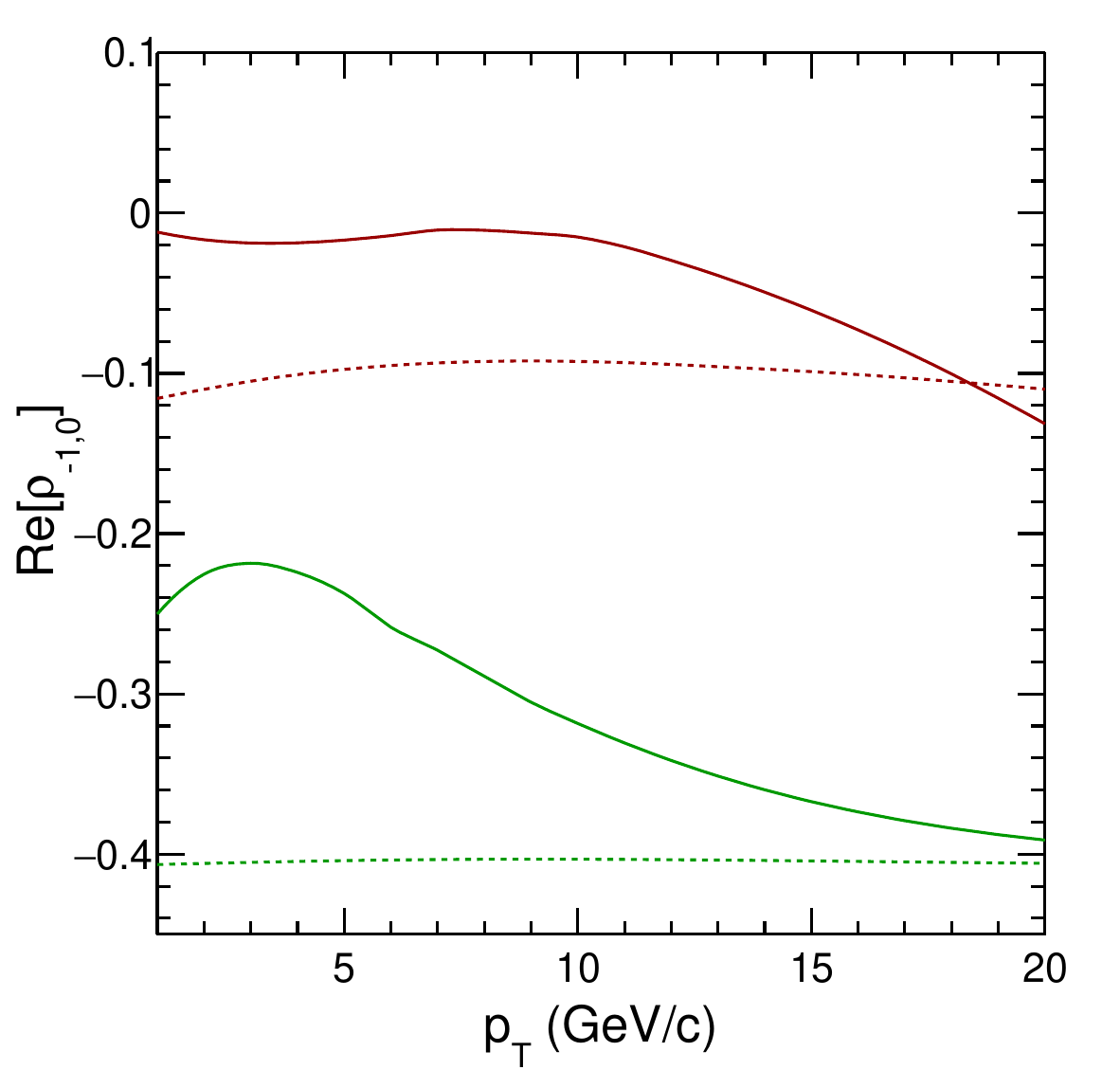}
\includegraphics[scale=0.29]{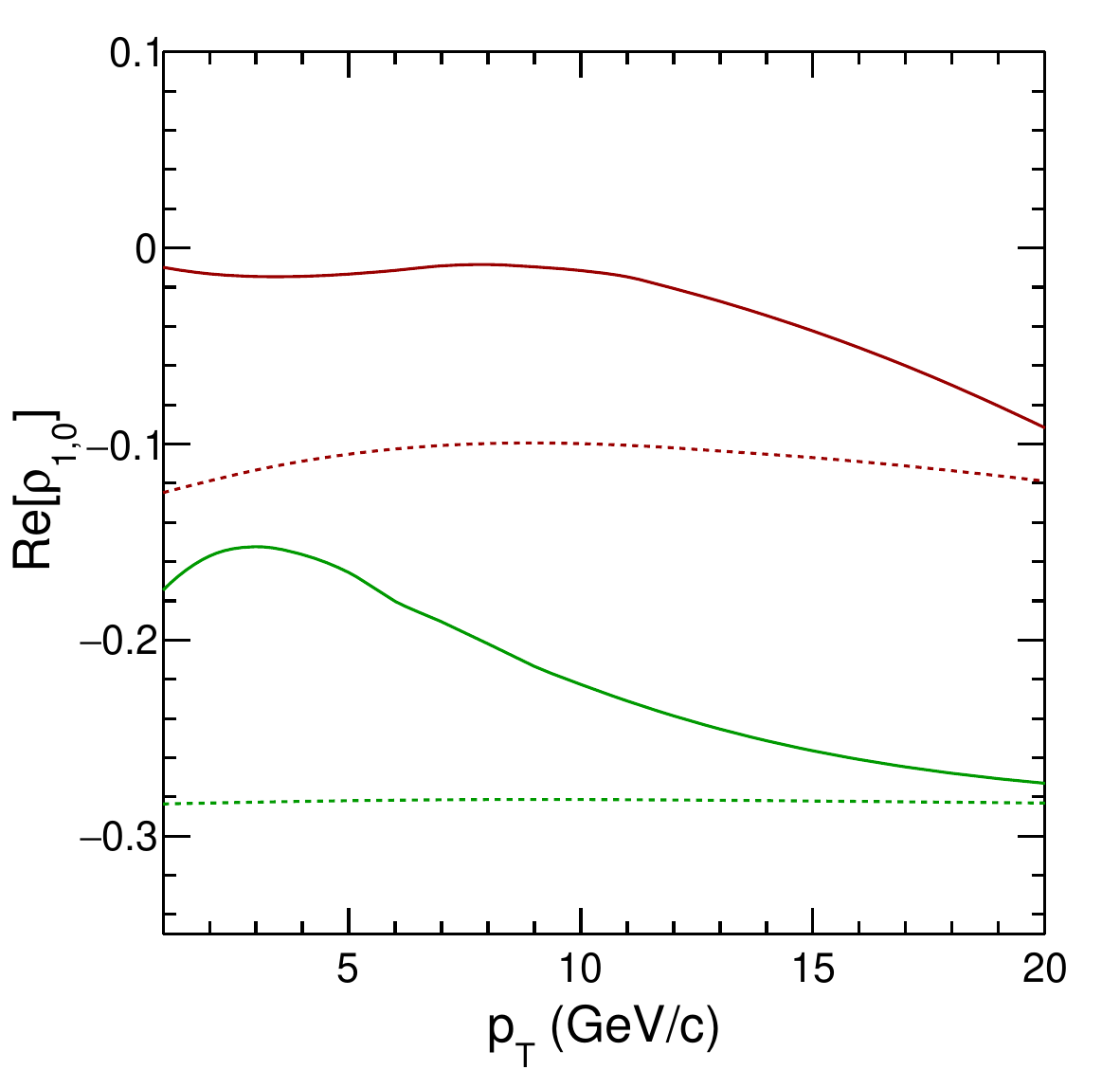}
\caption{(Color online) The off-diagonal elements of the spin density matrix, such as, Re[$\rho_{1,-1}$] (left panel), Re[$\rho_{-1,0}$] (middle panel), and Re[$\rho_{1,0}$] (right panel) as a functions of $p_{\rm T}$ for $J/\psi$, $\psi$(2S), $\Upsilon$(1S), and $\Upsilon$(2S) states for Pb+Pb collisions at $\sqrt{s_{\rm NN}}$ = 5.02 TeV calculated at mid-rapidity at $C=$ 1 fm.}
\label{fig:7}
\end{figure*}

\begin{figure*}[ht!] 
\includegraphics[scale=0.29]{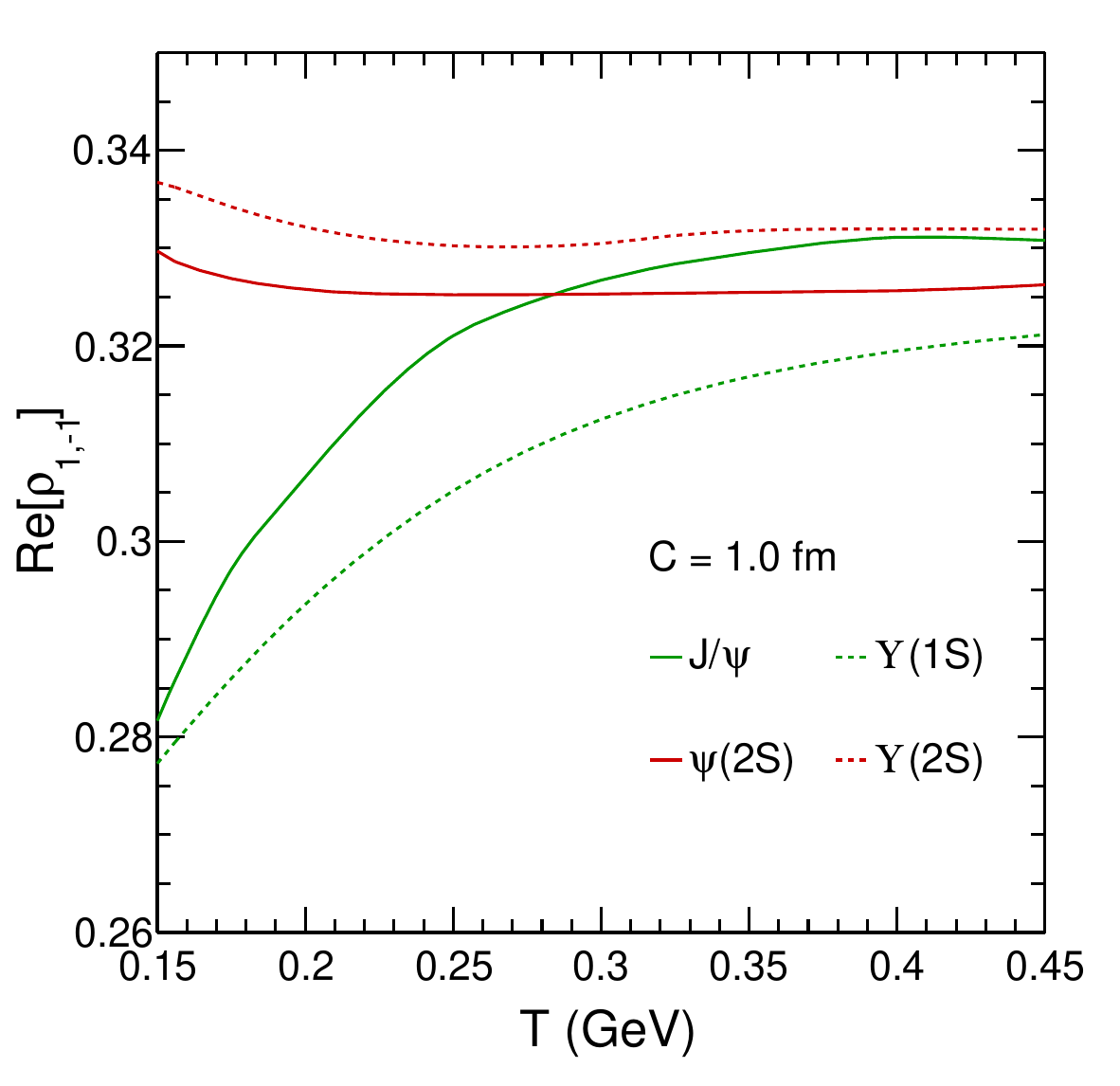}
\includegraphics[scale=0.29]{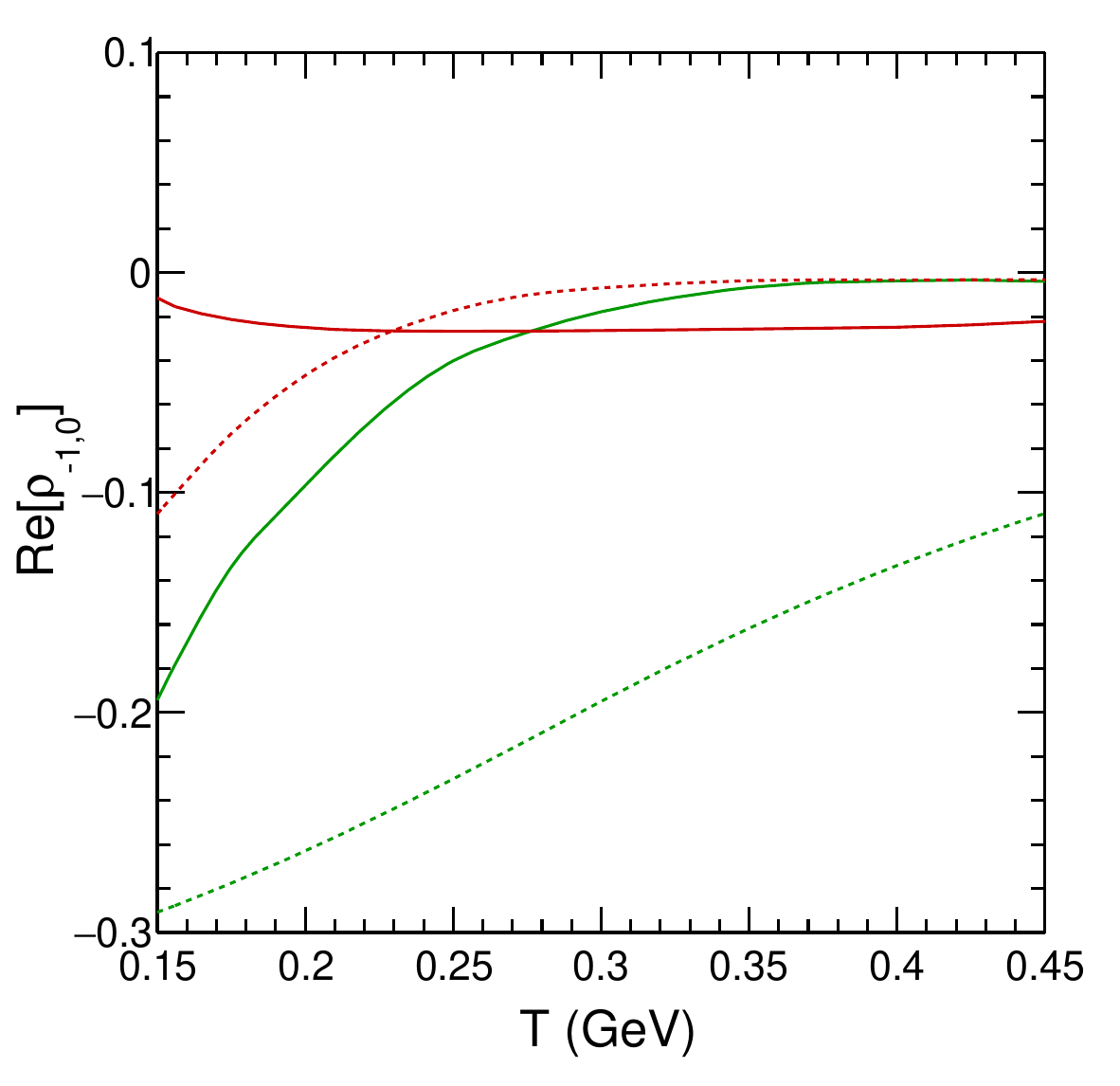}
\includegraphics[scale=0.29]{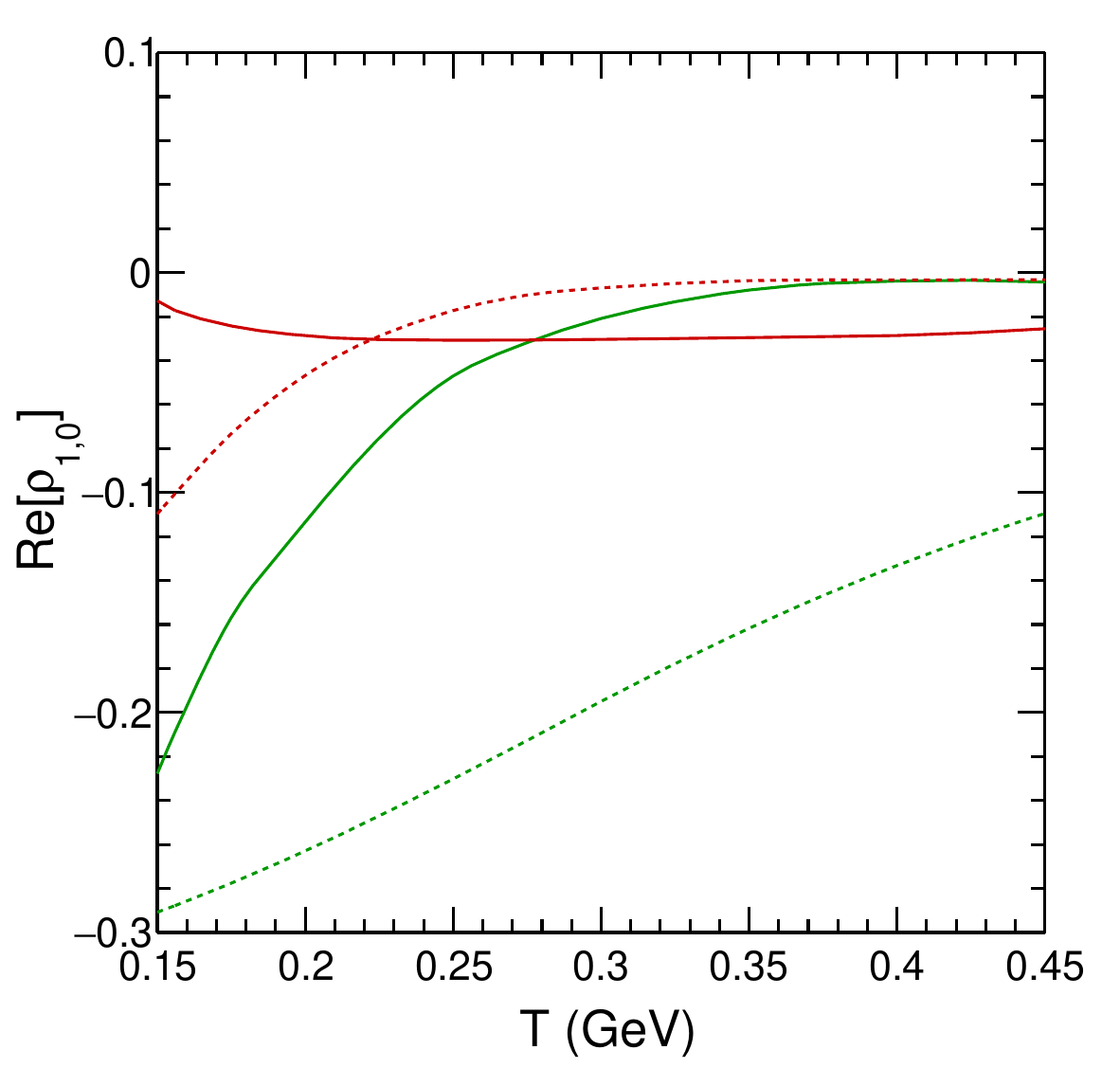}
\caption{(Color online) The off-diagonal elements of the spin density matrix, such as Re[$\rho_{1,-1}$] (left panel), Re[$\rho_{-1,0}$] (middle panel), and Re[$\rho_{1,0}$] (right panel) as a function of temperature $T$ for $J/\psi$, $\psi$(2S), $\Upsilon$(1S), and $\Upsilon$(2S) states at $C=$ 1 fm.}
\label{fig:8}
\end{figure*}

\begin{figure*}[ht!]
\includegraphics[scale=0.29]{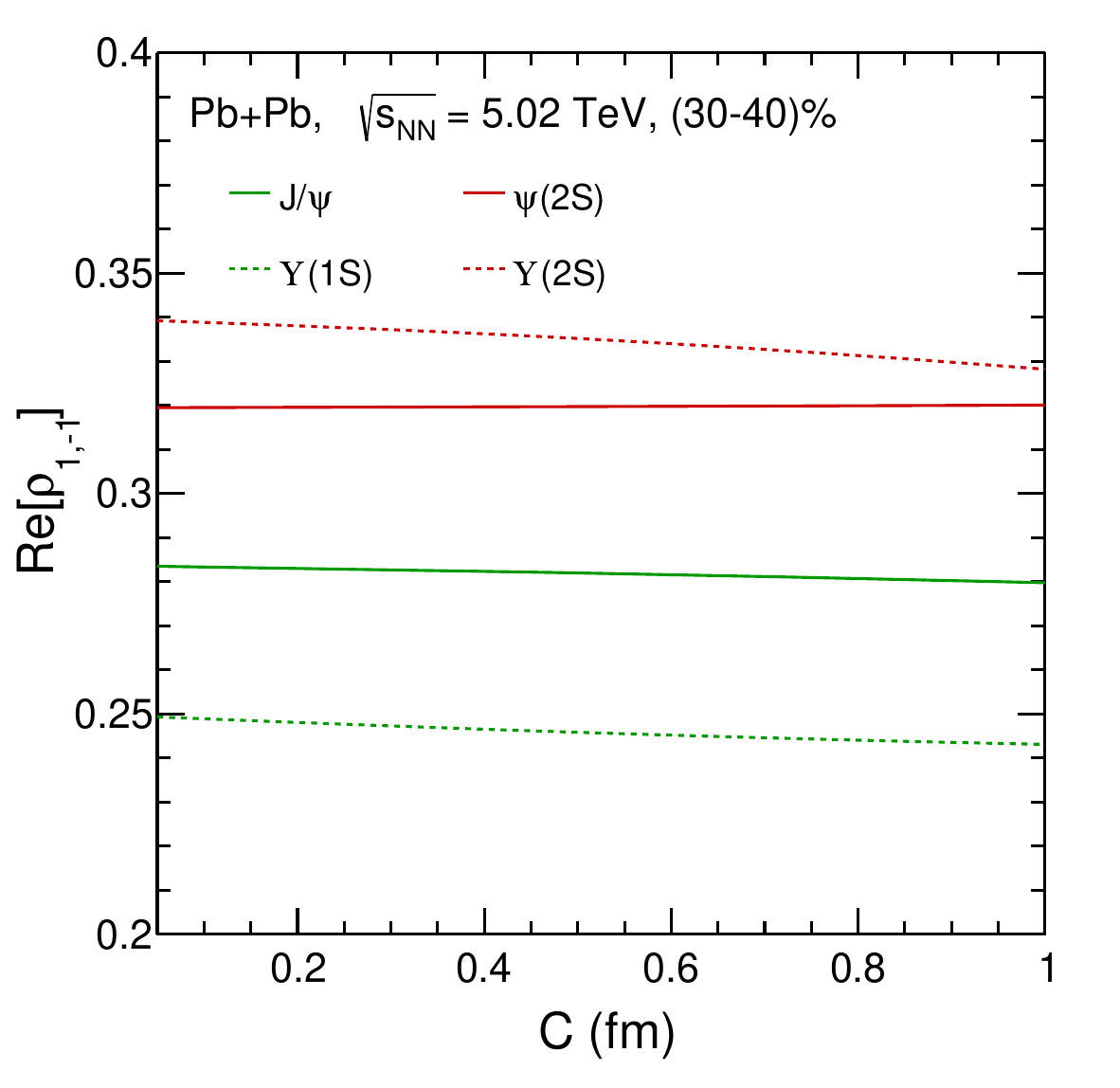}
\includegraphics[scale=0.29]{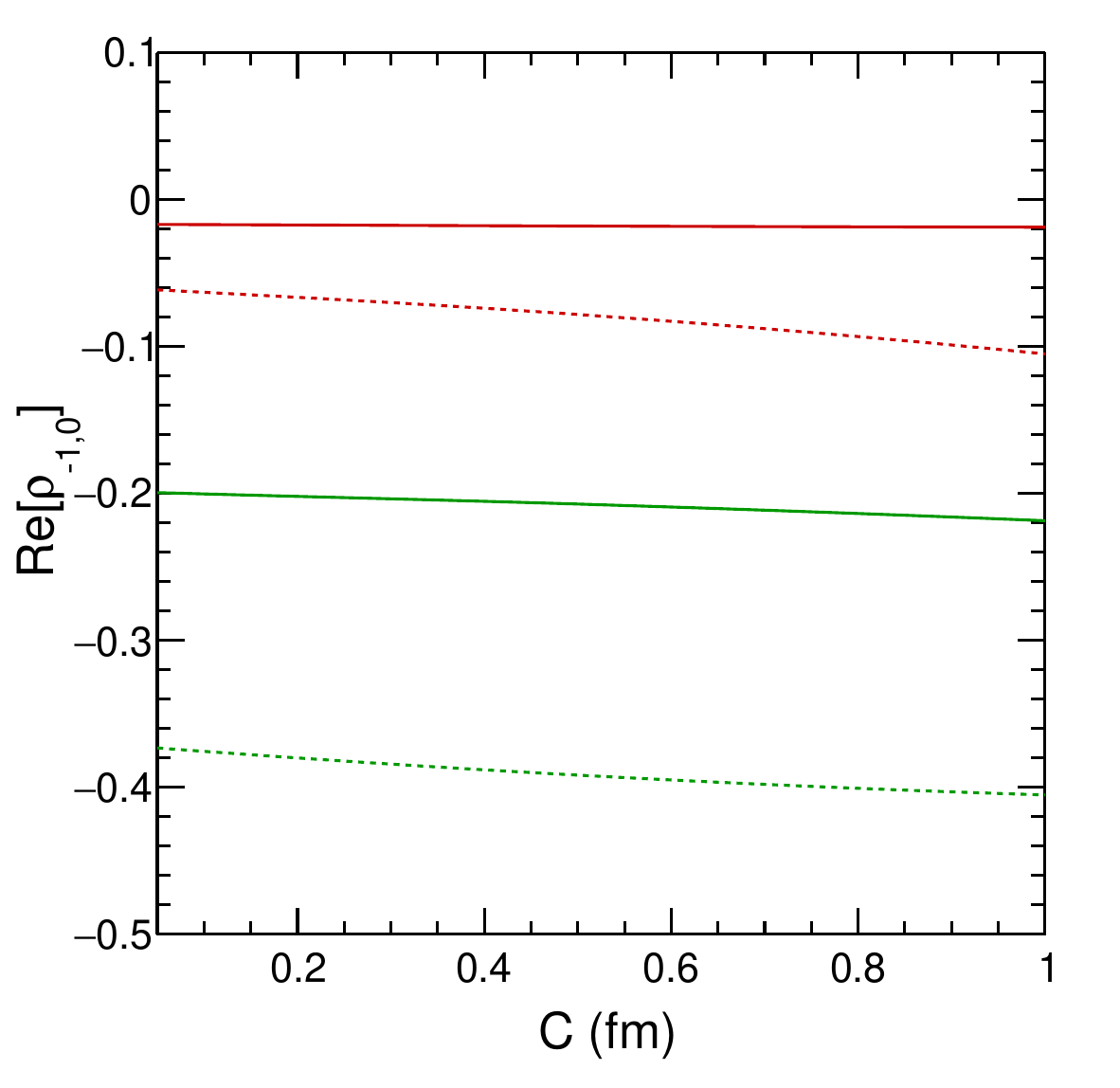}
\includegraphics[scale=0.29]{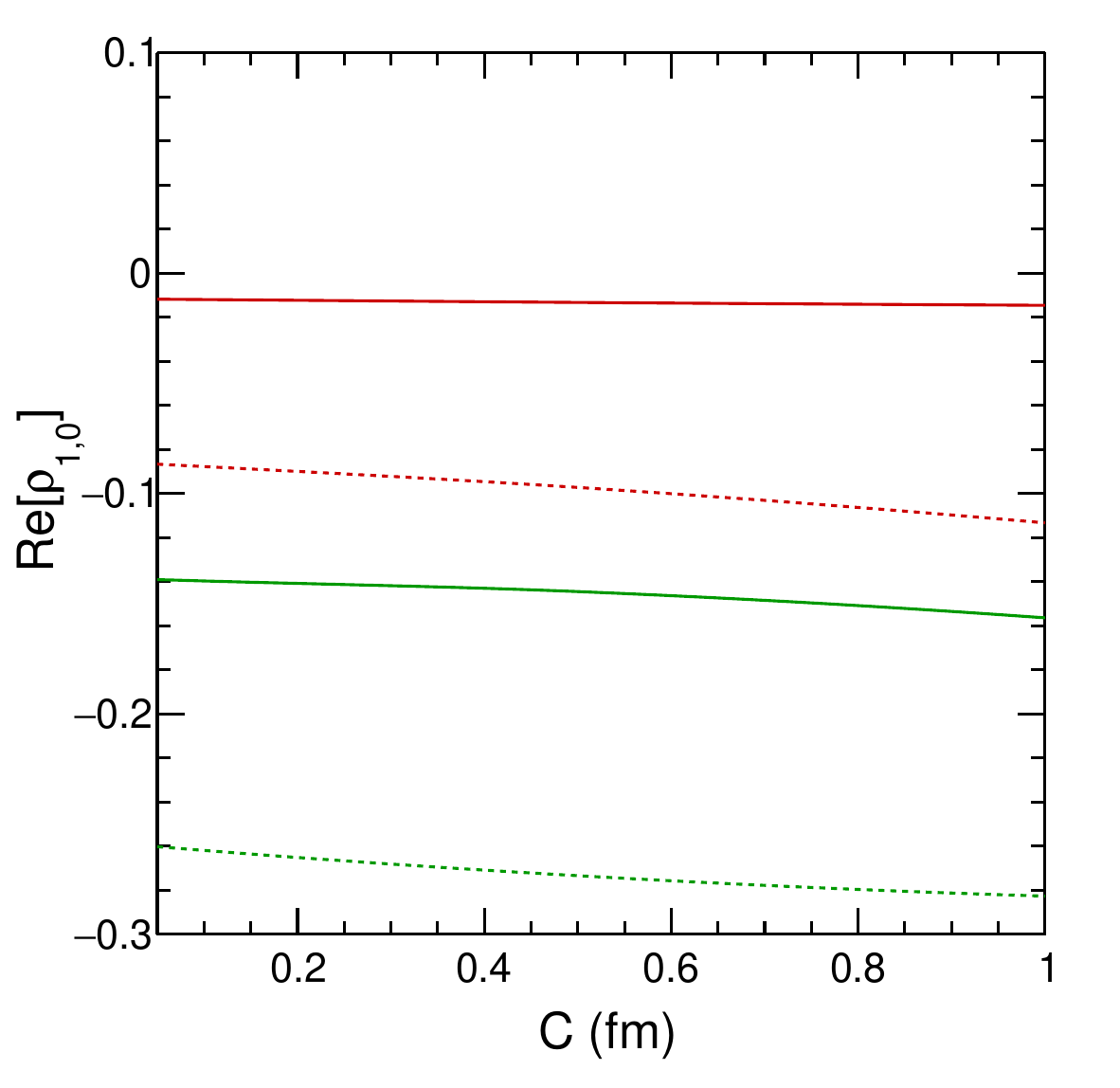}
\caption{(Color online) The off-diagonal elements of the spin density matrix, such as Re[$\rho_{1,-1}$] (left panel), Re[$\rho_{-1,0}$] (middle panel), and Re[$\rho_{1,0}$] (right panel) as a function of circulation parameter $C$ for $J/\psi$, $\psi$(2S), $\Upsilon$(1S), and $\Upsilon$(2S) states for Pb+Pb collisions at $\sqrt{s_{\rm NN}}$ = 5.02 TeV calculated at mid-rapidity.}
\label{fig:9}
\end{figure*}

\subsection{Off-diagonal elements}

Furthermore, it is proposed that apart from the diagonal elements of the spin density matrix $\rho_{00}$, the off-diagonal elements of the spin density matrix provide crucial information about the dynamics of the system produced in heavy-ion collisions. Such as the off-diagonal elements help to differentiate the local spin alignment of vector mesons from the global one. Studying these observables provides key insights about the spin hydrodynamic evolution in heavy-ion collisions and the freeze-out dynamics, etc~\cite{Xia:2020tyd, Goncalves:2021ziy, Goncalves:2024xzo, DeMoura:2023jzz}. It is proposed that the off-diagonal spin density matrix elements provide valuable insights into medium-induced spin correlations, quantum coherence, the local spin structure, and non-equilibrium QCD dynamics~\cite{Anselmino:1985bw}. Similar to diagonal elements, one can obtain the off-diagonal matrix elements of the spin$-$density matrix in a rotating frame, such as $\rho_{1, -1}$, $\rho_{-1, 0}$, and $\rho_{1, 0}$, and is expressed as follows;

\begin{enumerate}
\item $\rho^{r}_{1,-1}$
\begin{equation}
\rm \rho^{r}_{1,-1} = \frac{\sin^{2}{\theta_r}}{4} \left( \rho_{1,1} + e^{-2 i \phi_r} \rho_{-1,-1} + 2 e^{-i \phi_r} \rho_{0,0}\right)
\label{diagonalrho}
\end{equation}

At $\phi_r$ = 0, we have
\begin{equation}
\rm Re[ \rho^{r}_{1,-1}] = \frac{\sin^{2}{\theta_r}}{4} \left( \rho_{1,1} + \rho_{-1,-1} + 2 \rho_{0,0} \right)
\label{diagonalrho}
\end{equation}

Using the relation,  $\rho_{1,1} + \rho_{0,0} + \rho_{-1,-1} = 1$, we have 

\begin{equation}
\rm Re[ \rho^{r}_{1,-1}] = \frac{\sin^{2}{\theta_r}}{4} \left(1+\rho_{0,0}\right)
\label{diagonalrho}
\end{equation}

\item $\rho^{r}_{-1, 0}$

\begin{align}
\rm \rho^{r}_{-1,0} &= \frac{\sin \theta_r}{\;2 \sqrt{2}} \bigg[ (1- \cos \theta_r) \;  e^{2i \phi_r} \; \rho_{1,1} \nonumber \\ & + 2 \cos \theta_r e^{i \phi_r} \rho_{0,0} + (1+\cos \theta_r) \rho_{-1,-1} \bigg]
\label{diagonalrho}
\end{align}

At $\phi_r$ = 0, we have
\begin{align}
\rm Re[\rho^{r}_{-1,0}] &= \frac{\sin \theta_r}{\;2 \sqrt{2}} \bigg[ (1- \cos \theta_r) \; \rho_{1,1} \nonumber \\ & + 2 \cos \theta_r \rho_{0,0} - (1+\cos \theta_r) \rho_{-1,-1} \bigg]
\label{diagonalrho}
\end{align}

 \item $\rho^{r}_{1, 0}$

 \begin{align}
\rm \rho^{r}_{1,0} &= \frac{\sin \theta_r}{\;2 \sqrt{2}} \bigg[ (1+\cos \theta_r) \; \rho_{1,1} \nonumber \\ & - 2 \cos \theta_r e^{i \phi_r} \rho_{0,0} - (1-\cos \theta_r)\; e^{-2i \phi_r} \; \rho_{-1,-1}\bigg]
\label{diagonalrho}
\end{align}
At $\phi_r$ = 0, we have
 \begin{align}
\rm Re[\rho^{r}_{1,0}] &= \frac{\sin \theta_r}{\;2 \sqrt{2}} \bigg[ (1+\cos \theta_r) \; \rho_{1,1} \nonumber \\ & - 2 \cos \theta_r \rho_{0,0} - (1-\cos \theta_r)\; \rho_{-1,-1}  \bigg]
\label{diagonalrho}
\end{align}

\end{enumerate}

 Consequently, we explore the off-diagonal elements of the spin-density matrix, i.e., $\rho_{1,-1}$, $\rho_{-1,0}$, and $\rho_{1,0}$ for Pb+Pb collisions at $\sqrt{s_{\rm NN}}$ = 5.02 TeV, calculated at mid-rapidity in this section. The studied off-diagonal matrix elements appear as coefficients of the angular distribution of the decay daughter in the mother particle's rest frame. These coefficients are used as a proxy to measure the local polarization of the relativistic fluid. Figure.~\ref{fig:7} shows real part of the $\rho_{1, -1}$,  $\rho_{-1, 0}$, and $\rho_{1, 0}$ matrix elements as a function of $p_{\rm T}$ for $J/\psi$, $\psi$(2S), $\Upsilon$(1S), and $\Upsilon$(2S) states. The non-zero values of the off-diagonal elements in Fig.~\ref{fig:7} indicate that the system is in a mixed ensemble state and is characterized by the quantum superposition between the matrix elements.  As a result, the quantum coherence (decoherence) can be inferred from the spin-density matrix. On average, we found Re[$\rho_{1,-1}$],  Re[$\rho_{-1,0}$], and Re[$\rho_{1,0}$] have qualitatively similar trends as a function of $p_{\rm T}$, except a change in the magnitude is observed as a function of $p_{\rm T}$. It is observed that Re[$\rho_{1,-1}$] is always positive, while Re[$\rho_{-1,0}$] and Re[$\rho_{1,0}$] is always negative for all quarkonium states. Similar to $\rho_{00}$, $\Upsilon$(1S) and $\Upsilon$(2S) states have a very marginal effect on these off-diagonal matrix elements as a function of $p_{\rm T}$. However, for $J/\psi$ and $\psi$(2S), Re[$\rho_{1,-1}$],  Re[$\rho_{-1,0}$], and Re[$\rho_{1,0}$] show a decreasing trend at high $p_{\rm T}$.  The correlation between the different spin projections is reflected in these off-diagonal matrix elements.  It is proposed that both the local and long-range correlations can be estimated by studying the spin density matrix. Figure.~\ref{fig:8} and Fig.~\ref{fig:9} depict the real part of $\rho_{1, -1}$,  $\rho_{-1, 0}$, and $\rho_{1, 0}$ matrix elements as a function of temperature $T$ and circulation parameter $C$, respectively, for $J/\psi$, $\psi$(2S), $\Upsilon$(1S), and $\Upsilon$(2S) states.  Moreover, as presented in the study, the temperature and vorticity dependence of these off-diagonal elements reveal how the thermal and rotational characteristics of the medium impact the spin coherence. For example, we found $\rho_{-1,0}$ and $\rho_{1,0}$ approach zero with increasing temperature, reflecting a loss of quantum coherence at high temperature. Conversely, with increasing the circulation parameter $C$, which quantifies the vorticity of the medium, the magnitude of the off-diagonal elements for $\Upsilon$(1S) is enhanced (baseline is zero), whereas charmonium states as well as $\Upsilon$(2S) are marginally affected. This trend confirms that rotational effects promote quantum coherence between spin states, and hence, spin alignment is more pronounced in a thermo-vortical medium. 
 
\section{Summary}
\label{sum}

In this work, we have studied the spin alignment of $J/\psi$, $\psi$(2S), $\Upsilon$(1S), and $\Upsilon$(2S) states for Pb+Pb collisions at $\sqrt{s_{\rm NN}}$ = 5.02 TeV, calculated at mid-rapidity in a deconfined thermalized magneto-vortical medium in the presence of momentum-space anisotropy.  It is noteworthy to mention that the medium temperature is obtained using a Bjorken-like boost-invariant hydrodynamic formulation. However, a more realistic calculation can be performed using 3+1D hydrodynamics. The important findings of this paper are summarized below:

\begin{enumerate}

\item We explore the effect of vorticity, quantified in terms of the circulation parameter $C$, on the spin alignment of $J/\psi$, $\psi$(2S), $\Upsilon$(1S), and $\Upsilon$(2S) states by obtaining the vorticity and different spin projection-dependent energy eigenvalues. We use a medium-modified color-singlet potential model of quarkonia to estimate the medium temperature, vorticity, magnetic field, and medium momentum space anisotropy dependent on the energy eigenvalues by solving the Schr\"{o}dinger equation. Furthermore, the energy eigenvalues are used to calculate the diagonal element $\rho_{00}$, and the off-diagonal elements such as  $\rho_{1,-1}$, $\rho_{-1,0}$, and $\rho_{1,0}$ of the spin-density matrix for $J/\psi$, $\psi$(2S), $\Upsilon$(1S), and $\Upsilon$(2S) states in a rotating frame. 

\item The deviation of $\rho_{00}^{J/\psi}$ from 1/3 indicate that the distribution of the produced $J/\psi$ in relativistic heavy-ion collisions is not isotropic. In our study, $\rho_{00}^{J/\psi}$ is found to be always greater than 1/3 for all considered values of $C$ and $p_{\rm T}$ ranges. However, the recent experimental measurement shows $\rho_{00}$ is observed to be less than 1/3 at low $p_{\rm T}$ and greater than 1/3  at high $p_{\rm T}$ in Pb+Pb collisions at $\sqrt{s_{\rm NN}}$ = 5.02 TeV measured at forward-rapidity region~\cite{ALICE:2025cdf,ALICE:2022dyy,ALICE:2020iev}. 
Interestingly, similar to our findings of $\rho_{00}$ for $J/\psi$, and $\psi$(2S) shows increasing behavior of $\rho_{00}$ with $p_{\rm T}$  and large polarization is also observed for other heavy flavor hadrons such as $\rm D^{*+}$ in Pb+Pb collisions at $\sqrt{s_{\rm NN}}$ = 5.02 TeV\cite{ALICE:2025cdf, Dey:2025ail}. This hints at a common physics mechanism responsible for the spin alignment of heavy-flavor hadrons, and these heavy-flavor hadrons act as better probes for spin alignment studies in relativistic heavy-ion collisions. 

\item We found the $\rho_{00}$ observable for bottomonium states, specifically $\Upsilon$(1S) state, has pronounced effect ($\rho_{00} < $ 1/3 at low $C$ and $\rho_{00} >$ 1/3 at high $C$) of medium vorticity compared to the charmonium states.  A similar feature is also seen for the $\Upsilon$(2S) state as a function of $C$. This implies that the shape of the angular distribution depends strongly on the initial anisotropy that arises due to rotation. The value of vorticity, at which the shape of the angular distribution changes, is highly sensitive to the temperature and equilibration time of the quark-gluon plasma. Hence, the spin alignment of the bottomonium state may serve as a probe of system thermalization.

\item The effect of temperature, magnetic field, and medium momentum-space anisotropy on the $\rho_{00}$ observable is explored because the energy eigenvalues change with these parameters. The magnetic field and medium anisotropies affect the quarkonium states in a state-dependent manner. The possible reason for the state-dependent behavior of $ \rho_{00}$ in the presence of a magnetic field and medium anisotropy could be due to their different size, mass, binding energy, etc. Therefore, it is observed that the binding energy (mass) and melting temperature (width) of quarkonium states are sensitive to the medium temperature, rotation, magnetic field, and anisotropy parameters. In addition, it is expected that the coupling strength and the time scale at which the bottom and charm quarks will align with the magnetic field and the anisotropy direction will be different. This may affect the spin alignment observables of different quarkonium states. 

\item The non-vanishing value of the off-diagonal elements of the spin-density matrix, such as $\rho_{1,-1}$, $\rho_{-1,0}$, and $\rho_{1,0}$, reflects the possible coherence between different spin states. This could be due to the final-state interactions of the vector mesons. The $\rho_{00}$ determines the magnitude of spin alignment along a quantization axis, and along with the off-diagonal components, the spin density matrix describes the full three-dimensional structure of the spin distribution, including angular correlations and coherence effects. Understanding and measuring these elements experimentally can thus unlock deeper insights into the collective behavior of the QGP and the fundamental mechanisms of spin transport in strongly interacting matter, paving the way for future advancements.

\item It is important to note that apart from the studied sources, there could be various potential sources and physics mechanisms responsible for the spin-alignment of quarkonium states. Such as the correlations and fluctuations of the strong force field between the quark and anti-quark pairs~\cite{Sheng:2022wsy}, turbulent color field~\cite{Kumar:2023ghs, Muller:2021hpe}, electric field~\cite{Sheng:2019kmk, Yang:2017sdk}, fragmentation process~\cite{Liang:2004xn}, local spin alignment and helicity~\cite{Xia:2020tyd, Gao:2021rom}, etc. The contribution of each source to the spin alignment of quarkonium states requires more examination. \\\ 
\end{enumerate}

Apart from the degree of alignment of different hadrons, the spin alignment measurement in heavy-ion collisions provides various other crucial features of the system. Such as the $\rho_{00}$ elements affect the experimentally measurable chiral magnetic effect (CME) sensitive observables, including $\gamma_{112}$ correlator, $R_{\Psi_2}(\Delta S)$ correlator, and signed balance function, etc.~\cite{Shen:2022gtl}. Furthermore, the $\rho_{00}$ observable helps to find the correlations and fluctuations of the strong force field in the quarkonium states as it is linked to the fundamental properties of QCD. Quarkonium polarization could be a promising observable to probe the spin hydrodynamics and freeze-out properties in the quark-gluon plasma. So far, the spin alignment of various vector mesons across collision systems and energies remains challenging and not described by a common physics mechanism. Thus, an intense investigation of the spin alignment of vector mesons is required from both experimental and theoretical perspectives.
With the current high statistics, the Run 3 data samples of
LHC experiments, a more differential and precise measurement
can be performed on the local and global spin alignment
of various vector mesons (including strange, charm, and bottom sectors). 

\section*{Acknowledgement}
Bhagyarathi Sahoo acknowledges the financial aid from CSIR, Government of India. The authors gratefully acknowledge the DAE-DST, Government of India, funding under the mega-science project "Indian Participation in the ALICE experiment at CERN" bearing Project No. SR/MF/PS-02/2021-IITI (E-37123).

\end{document}